\documentclass[12pt]{iopart}

\usepackage{graphicx}
\usepackage[toc]{appendix}
\usepackage{color}
\usepackage{url}
\usepackage{amsfonts}
\usepackage{amssymb}
\usepackage{dcolumn}
\usepackage{subfig}
\usepackage{verbatim}
\usepackage{qcircuit}
\usepackage{tikz,pgfplots}
\usepackage[hidelinks]{hyperref}

\def\plog{\overline{P}}

\newcommand{\ket}[1]{| #1 \rangle}

\def\t{T}

\begin{document}

\title{Hyperbolic and Semi-Hyperbolic Surface Codes for Quantum Storage}
\author{Nikolas P. Breuckmann$^1$, Christophe Vuillot$^1$, Earl Campbell$^2$, Anirudh Krishna$^3$ and Barbara M. Terhal$^1$}
\address{$^1$ JARA Institute for Quantum Information, RWTH Aachen University, Aachen, Germany\\
$^2$ Department of Physics and Astronomy, University of Sheffield,
Sheffield, UK\\
$^3$ Departement de Physique, Universite de Sherbrooke, Canada}
\eads{\mailto{nikobreu@gmail.com}, \mailto{vuillot@physik.rwth-aachen.de}, \mailto{earltcampbell@gmail.com}, \mailto{krishnanirudhster@gmail.com}, \mailto{bterhal@gmail.com}}

\date{\today}
\begin{abstract}
We show how a hyperbolic surface code could be used for overhead-efficient quantum storage. 
We give numerical evidence for a noise threshold of $1.3\%$ for the $\{4,5\}$-hyperbolic surface code in a phenomenological noise model (as compared to $2.9\%$ for the toric code). 
In this code family parity checks are of weight 4 and 5 while each qubit participates in 4 different parity checks. We introduce a family of semi-hyperbolic codes which interpolate between the toric code and the $\{4,5\}$-hyperbolic surface code in terms of encoding rate and threshold. We 
show how these hyperbolic codes outperform the toric code in terms of qubit overhead for a target logical error probability. We show how Dehn twists and lattice code surgery can be used to read and write individual qubits to this quantum storage medium.
\end{abstract}

\section{Introduction}

The surface code has become a preferred coding architecture due to its high noise threshold and its relatively simple use of 2D connectivity between qubits. 
For the surface code, both parity check weight and qubit degree (meaning the number of parity checks that a qubit participates in) are low, namely 4. 
Together with code deformation to perform Clifford gates and techniques such as magic state distillation, the surface code could be used as a coding platform for universal computation, see e.g. \cite{terhal:rmp, CTV:review} and references therein. 

One disadvantage of the surface code is its spatial overhead: the number of physical qubits per logical qubit that allow one to achieve a certain logical error probability. 
For classical or quantum storage the use of block codes which encode $k$ logical qubits into $n$ physical qubits  can lead to qubit overhead savings.  The use of block codes in quantum computation was previously considered in \cite{SI:block} and \cite{brun+:tele_block}.
Since two-dimensional topological quantum codes based on Euclidean tilings have to obey the Bravyi-Terhal-Poulin bound $kd^2 \leq c n$ for some constant $c$, the savings of such block codes are limited (see e.g. \cite{DIP:overhead}). 
Hyperbolic surface codes based on tilings of a closed hyperbolic surface are not limited by this bound. They are families of codes with an asymptotically constant rate $k/n \geq c_1$ while the distance $d$ of these codes can be lower bounded by $c_2 \log n$ for some constants $c_1$ and $c_2$ \cite{BT:hyper}.
Delfosse \cite{D:tradeoffs} has proved that this logarithmic scaling of the distance is the best one can get for codes based on closed two-dimensional surfaces: the general trade-off bound is $k d^{2} \leq c (\log k)^{2} n$ with a constant $c$. 

In previous work \cite{BT:hyper} we have shown how to construct families of hyperbolic surface codes and we have presented numerical evidence for their noise threshold with respect to Pauli noise and the encoding overhead for a given logical error probability in case of noise-free parity check measurements. 
In \cite{DIP:bench} the authors numerically determine thresholds against erasure errors for large hyperbolic codes.
In this work we focus on a particular promising family of hyperbolic surface codes, namely codes based on a $\{4,5\}$-tiling and extend the results to noisy parity checks.  
We exhibit some of the large overhead savings that one may get from using these codes. 

The price to pay for these codes is that they require a connectivity between physical qubits that is not geometrically local in 2D Euclidean space.
However, 2D connectivity is not a necessity for some qubit implementations such as trapped ions or NV-centers linked via photonic couplings \cite{vandam+:NV, monroe+:photonic, KMNRDM:photonic_qc,NFB:photonic_links} or silicon photonic architectures \cite{GSBR:ballistic}. 
For these architectures a reasonable scalability criterion would be to demand a low, constant connectivity between qubits.

In the remainder of this section we will review hyperbolic surface codes and the modification to what we call {\em semi-hyperbolic} surface codes. The class of semi-hyperbolic surface codes allows one to interpolate between the logarithmic distance scaling $\log n$ of the hyperbolic codes 
and the $\sqrt{n}$ scaling of the surface code as well as interpolate between the thresholds of these code families.

In Section \ref{sec:num} we then consider the performance of these codes in the presence of both qubit and measurement errors.  
In previous work \cite{BT:hyper} we considered noiseless parity checks and found a considerably worse threshold for hyperbolic codes as compared to the surface code. One of the interesting findings of this paper is that for noisy parity checks the performance of hyperbolic surface
codes is much more comparable to the surface code itself. 
In Section \ref{sec:noise} we analyze the logical error probability borrowing some ideas from \cite{dklp}. 

In Section \ref{sec:overhead} we quantify how much overhead savings are possible. We do this numerically and semi-analytically: we provide an approximate formula for the logical error probability which works well in the low error rate regime.

One important question to address is how one can compute on individual qubits stored in a block code. In Section \ref{sec:logic} we present our ideas on how can read or write qubits from storage, using Dehn twists
to move qubits around in storage. Our proposed methods keep constant connectivity between qubits at all steps, as well retaining the qubit overhead savings.
We close the paper with a discussion and open questions on embedding hyperbolic surface codes in a Euclidean bilayer or 3D space and further exploration of these codes.

\subsection{Hyperbolic Surface Codes}
\label{sec:HSC}

Hyperbolic surface codes are, like the toric code, examples of homological CSS codes \cite{freedman_projective, freedman2002z2}. 
In principle one can obtain a CSS surface code by gluing polygons together at their edges to form a surface.  One puts qubits on the edges of these polygons and associates $Z$-checks with each polygon face, acting on the edges of the polygon. 
One associates an $X$-check with each vertex where some polygons meet such that the $X$-check acts on all edges emanating from the vertex. 
In order to encode logical information, one needs to either (1) punch holes in the created surface, or (2) create boundaries on which logical operators can terminate or (3) identify polygons to make the whole object a closed surface with handles.
But such a construction is not a systematic way to obtain code families with certain guaranteed properties in terms of distance or rate, while 
hyperbolic surface codes do offer a systematic construction. A family of hyperbolic surface codes can be obtained as follows (see \ref{app:tiling} for more details).
First, one chooses a regular tiling of the hyperbolic plane, given by its Schl\"afli symbol $\{r,s\}$. 
The Schl\"afli symbol indicates that $r$ regular $s$-gons meet at each vertex of the tiling. 
The only regular tilings that give a flat surface are $\{4,4\}$ (square lattice) and $\{6,3\}$ (hexagonal lattice) or its dual  $\{3,6\}$ (triangular lattice). 
If one chooses any $r$ and $s$ for which $1/r + 1/s < 1/2$ one obtains a tiling of hyperbolic space with negative curvature.
In this paper we will focus on the $\{4,5\}$-tiling, meaning that five squares meet at each vertex, as it has more favorable properties \cite{BT:hyper} and is the most similar to the toric code. 
In order to encode qubits the tiled surface has to be topologically non-trivial. We will only consider the case where we have a tiled, closed surface with many handles.  
For such a surface with edges $E$, faces $F$ and vertices $V$, the number of edges $n=|E|$ equals the total number of qubits. 
The number of parity $Z$-checks equals $|F|$ and each parity $Z$-check is of weight $r$. The number of parity $X$-checks equals $|V|$ and each parity $X$-check, corresponding to a vertex, acts on $s$ qubits. Such a code encodes $k=2g$ logical qubits, where $g$ is the number of handles (genus) of the surface.
The parameters $[[n,k,d]]$ of the code are determined by how the hyperbolic surface is closed. 
In \ref{app:tiling} we review how this closing procedure works. More details can be found in \cite{BT:hyper}.

As a concrete example one can consider a $\{3,7\}$-tiling of a genus 3 surface called the Klein quartic.
The Klein quartic quantum code has parameters $[[84, 6, 4]]$ and an embedding of the surface in 3D is shown at \url{http://math.ucr.edu/home/baez/klein.html}. In Figure \ref{fig:KQbilayer} in the Discussion we show an explicit embedding of this tiling in a two-dimensional bilayer.
In \cite{isaac} (see also \cite{vogeler}) the distances of hyperbolic codes based on $\{3,7\}$-tilings were studied.

\begin{figure}[htb]
\centering
    \subfloat[Tiling of a genus 4 surface with the $\{5,4\}$-lattice (which is dual to the $\{4,5\}$-lattice). Edges with the same label are identified. For the code that we analyze the $X$-checks are given by faces and the $Z$-checks by vertices. A $\overline{X}$ of weight 6 and a $\overline{Z}$ of length 4 are highlighted.]{
      \includegraphics[width=0.8\linewidth]{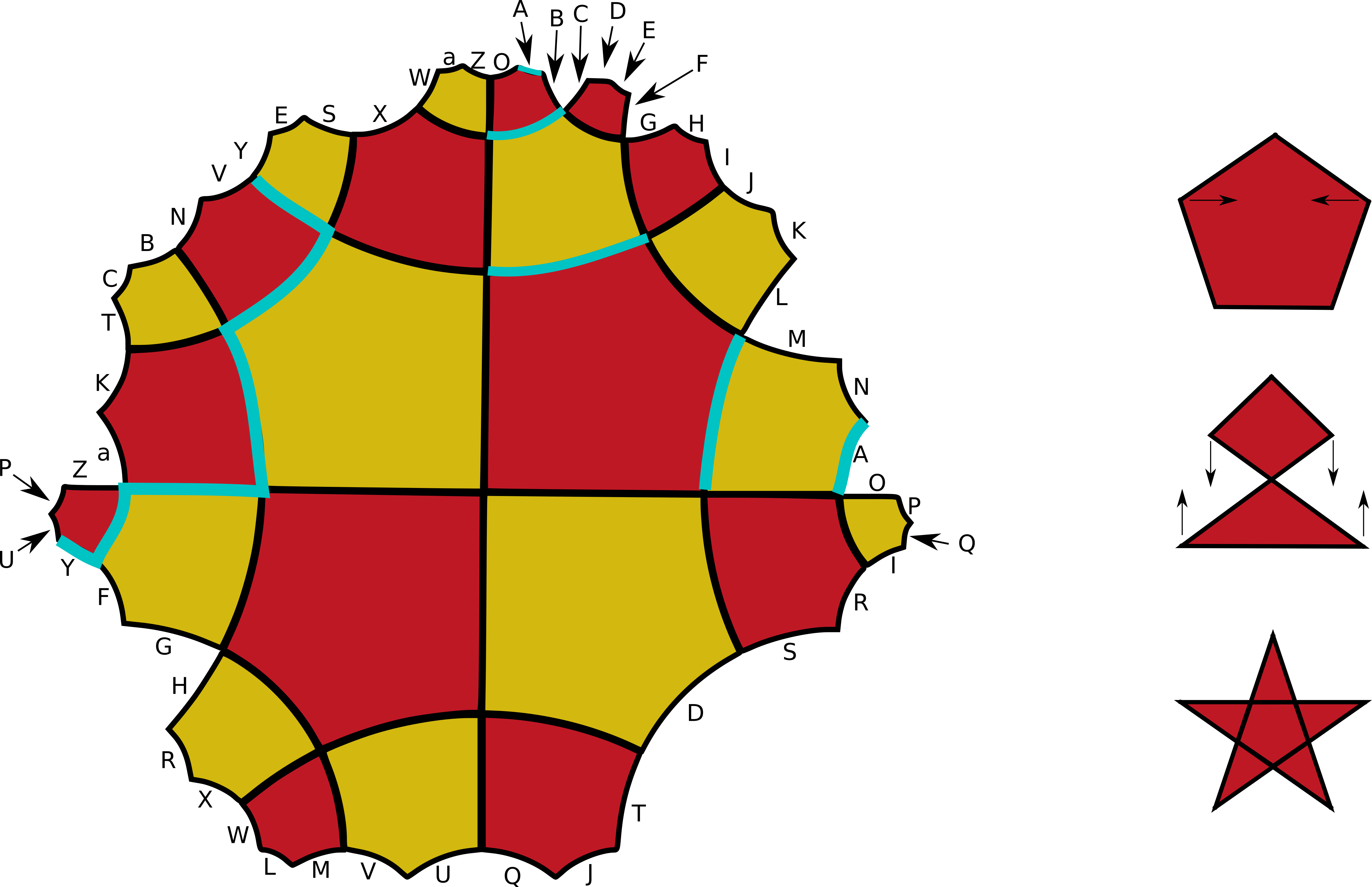}
    }

        \subfloat[Weight-6 $\overline{X}$.]{
        \includegraphics[width=0.3\linewidth]{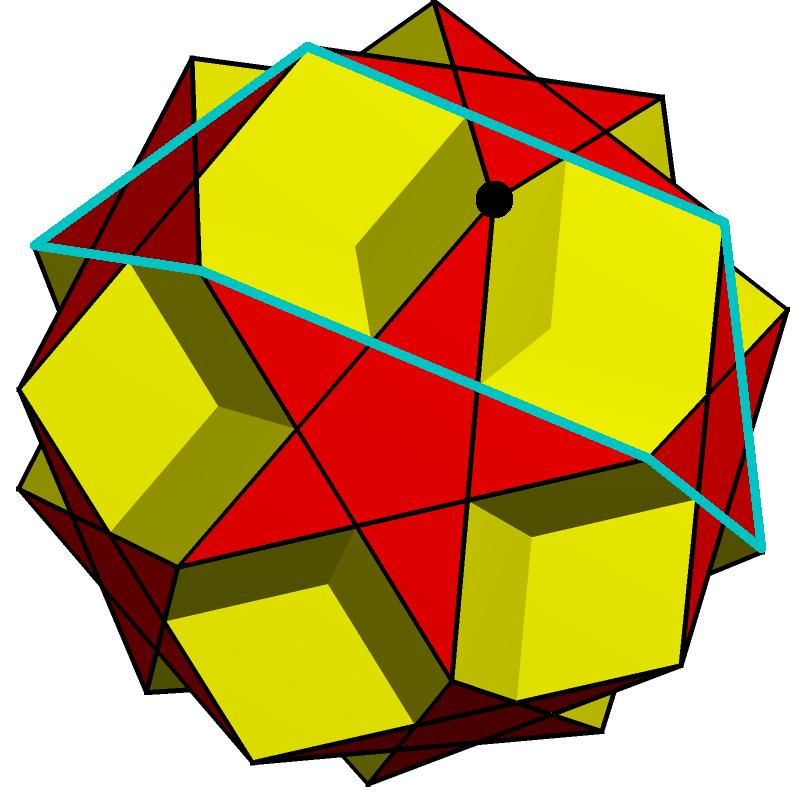}
    }\hspace{0.2cm}
    \subfloat[Weight-6 $\overline{X}$.]{
        \includegraphics[width=0.3\linewidth]{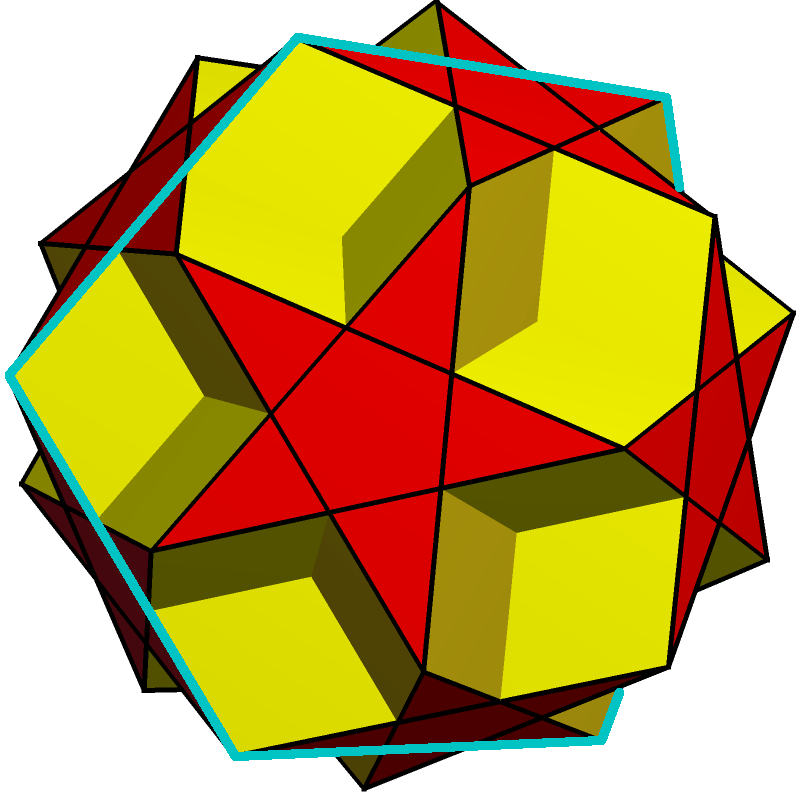}
    }\hspace{0.2cm}
    \subfloat[Weight-4 $\overline{Z}$.]{
        \includegraphics[width=0.3\linewidth]{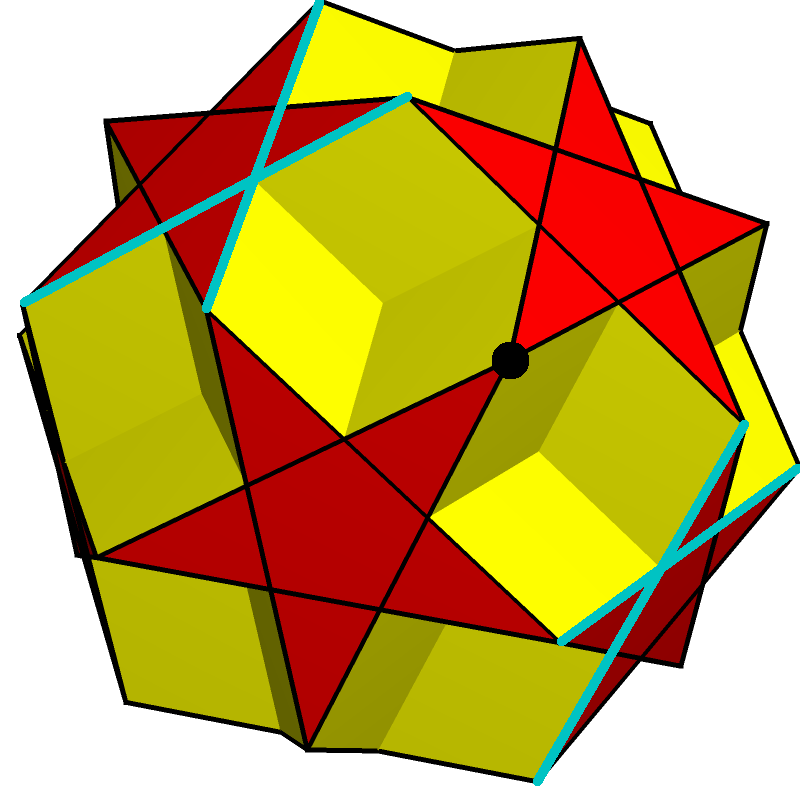}
    }
    \caption{The smallest code based on a $\{4,5\}$-tiling has parameters $[[60,8,4]]$ and is related to the so called {\em dodecadodecahedron}. This can be seen as follows: We use the dual $\{5,4\}$-tiling where four pentagons meet at every vertex. Half of the faces (red) are deformed into pentagrams with self-intersecting edges, see (a). Arranging the vertices on the surface of a sphere and allowing for self-intersecting faces, gives the dodecadodecahedron, see (b-d). A vertex is highlighted by a dot in order to show that one can label some of the logical operators by the 30 vertices (see \ref{sec:dist_comput}).}
\label{fig:dodecadodecahedron}
\end{figure}

For the $\{4,5\}$-tiling the smallest four examples are $[[60,8,4]]$, $[[160,18,6]]$, $[[360,38,8]]$ and $[[1800,182,10]]$ with an asymptotic rate $k/n \rightarrow 1/10$ (see Table \ref{tab:codes}). In Figure \ref{fig:dodecadodecahedron} we show the smallest code $[[60,8,4]]$ with respect to the dual tiling $\{5,4\}$. This code has a nice representation in 3D as a self-intersecting star polyhedron. Other small quantum codes could be constructed from such polyhedra with non-trivial genus (see listings on Wikipedia). The distance of these various small codes based on their corresponding surfaces can be computed with the method in \ref{sec:dist_comput}. The polyhedral representation immediately suggests a 3D qubit layout.

\subsection{Semi-Hyperbolic Surface Codes}
\label{sec:semi}

\begin{figure}[htb]
\centering
    \includegraphics[width=0.4\linewidth]{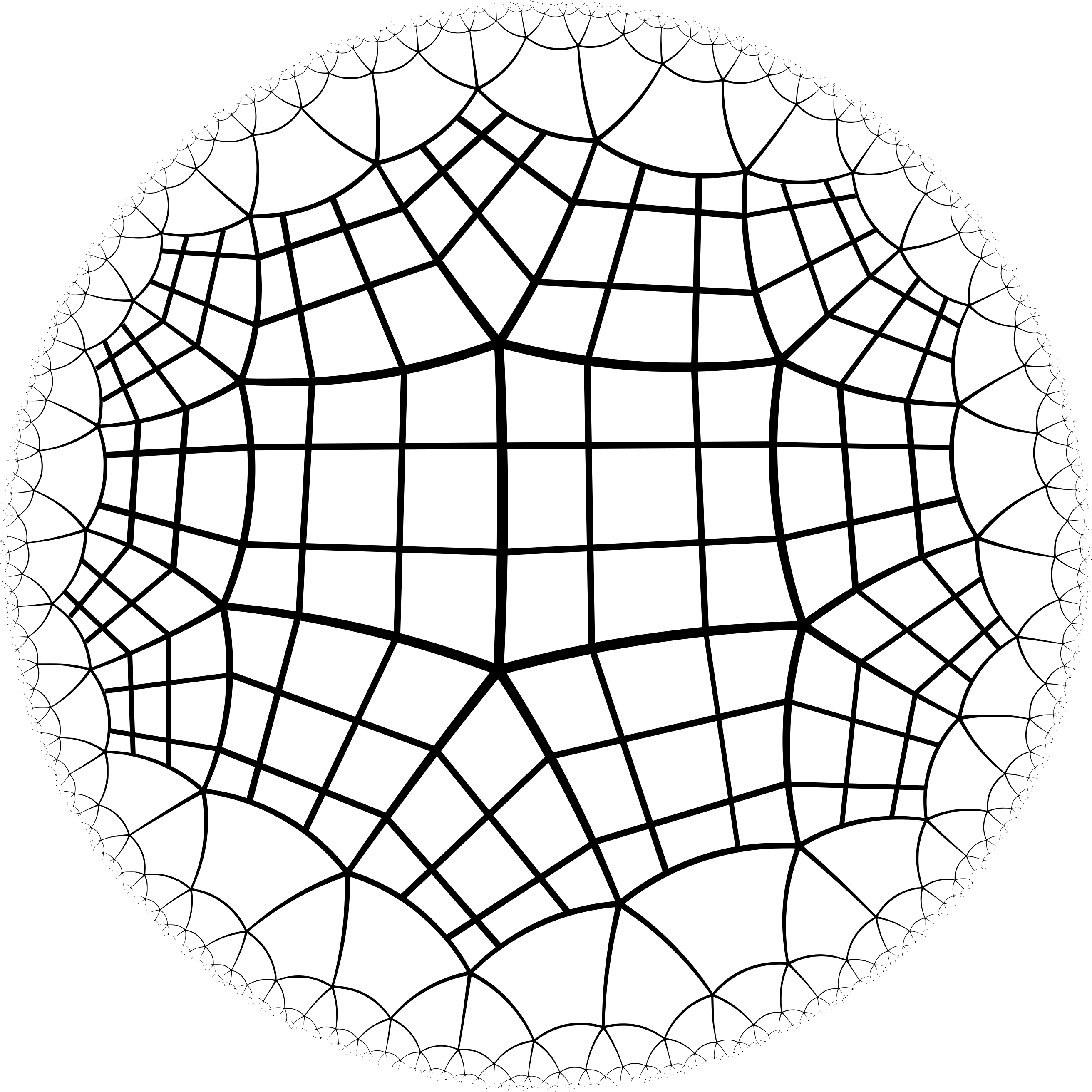}
    \caption{The $\{4,5\}$-lattice with some faces replaced by a $3\times 3$ square grid.}
    \label{fig:semi-hyp_lattice}
\end{figure}

Consider a regular tiling of a closed, hyperbolic surface with Schl\"afli-symbol $\{4,q\}$ ($q\geq 5$), given by a set of vertices $V$, edges $E$ and faces $F$. 
In a $\{4,q\}$-tiling it holds that $|E| = q|V|/2 = 2|F|$.  Associated with this tiling is a $\{4,q\}$-hyperbolic surface code with parameters $[[n_h,k_h,d_h]]$.
We define a new tiling with the same topology by taking every face and tiling it with an $l\times l$ square-grid lattice (see Figure \ref{fig:semi-hyp_lattice}). 
Essentially we replace each square by a $\{4,4\}$-tiling of a 2D flat space, weakening the negative curvature.

\begin{figure}
    \centering
    \includegraphics{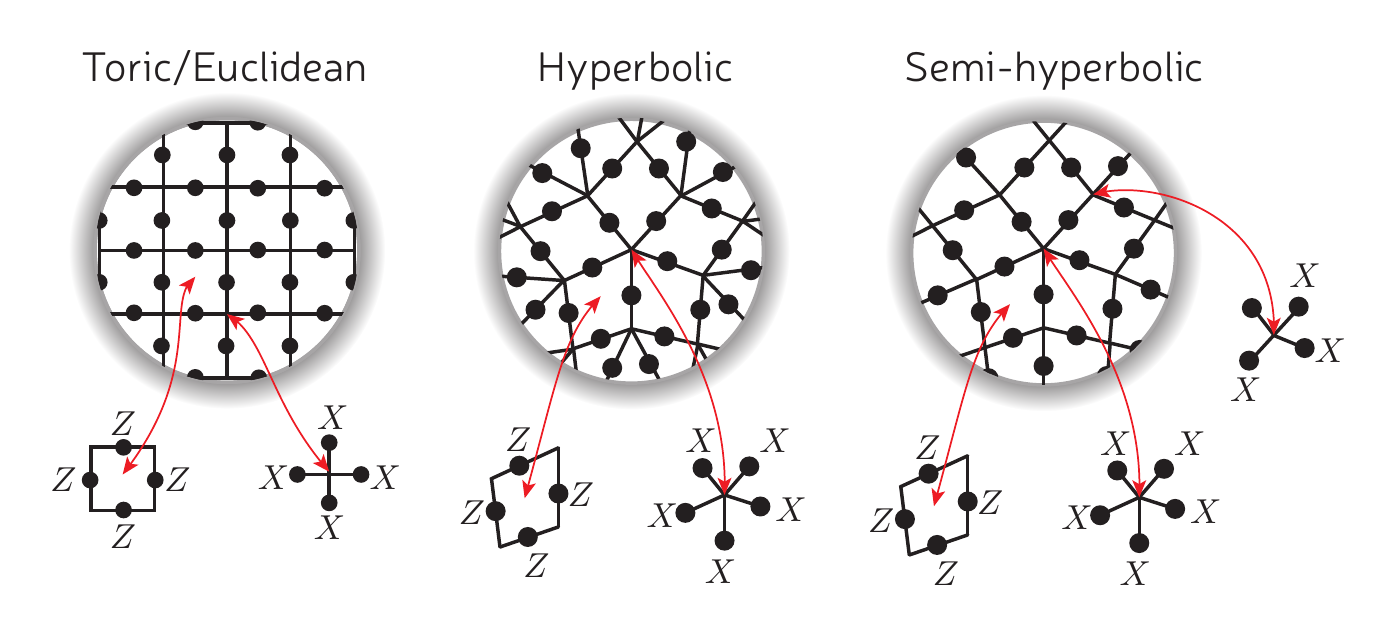}
    \caption{Local regions in a $\{4,4\}$-lattice (left), $\{4,5\}$-hyperbolic lattice (middle) and a semi-hyperbolic lattice based on the $\{4,5\}$-lattice (right).}
    \label{fig:local_regions}
\end{figure}

We call this new lattice \emph{semi-hyperbolic} having vertices $V_{sh}$, edges $E_{sh}$ and faces $F_{sh}$ with
\begin{eqnarray}\label{eqn:semihyp_lattice_params}
        |V_{sh}| &= |F| l^2 = q |V| l^2 / 4,\nonumber\\         |E_{sh}| &= |F|\times  2l^2 = |E| l^2,\\  |F_{sh}| &= |F| l^2.\nonumber
\end{eqnarray}
 From (\ref{eqn:semihyp_lattice_params}) it immediately follows that $n_{sh} = n_h l^2$. The number of encoded qubits in the hyperbolic surface code is determined by the topology of the surface which is unchanged, hence $k_{sh} = k_h$. 
For semi-hyperbolic codes, all $Z$-checks have weight 4 while there are two types of $X$-checks, namely the ones of weight $q$ of the original code and the new checks of weight 4 of which there are $|V| (q l^2 / 4 - 1)$ (see Figure \ref{fig:local_regions}).  
One can efficiently compute (in ${\rm poly}(n)$ steps) the distance of $\overline{Z}$ and $\overline{X}$ for CSS surface codes \cite{bravyi_algo}. We review this algorithm to compute distances and its uses in \ref{sec:dist_comput}. 
Our results are listed in Table \ref{tab:codes}. They support the conjecture that the $\overline{Z}$-distance of the semi-hyperbolic code is $d_{sh}(\overline{Z}) =d_h l$.
This would be true if the shortest non-trivial loops go over the subdivided squares {\em through} the vertices of the original hyperbolic code lattice. We have not been able to prove this however. Table \ref{tab:codes} shows that the scaling of the $\overline{X}$-distance is 
clearly also growing with $l$ although the $l$-dependence is not as simple as the conjectured $l$-dependence of the $\overline{Z}$-distance.

With increasing $l$ the ratio of total curvature over the surface area vanishes so one expects that for fixed $n_h$ and increasing $l$ a semi-hyperbolic code family has similar behavior to the toric code in terms of noise threshold. 
We confirm this in Figure \ref{fig:semi-hyp_threshold} in Section \ref{sec:overhead}.

\begin{table}[htb]
	\begin{center}
		\begin{tabular}{|c|c|c|c|c|c|}
        \hline $n_h$ & $l$ & $n$ & $k$ & $d(\overline{Z})$ & $d(\overline{X})$ \\
        \hline 60 & 1 & 60 & 8 & 4 & 6 \\
      \hline 60 & 2 & 240 & 8 & 8 & 10 \\
      \hline 60 & 3 & 540 & 8 & 12 & 14 \\
      \hline 60 & 4 & 960 & 8 & 16 & 18 \\
      \hline 60 & 5 & 1500 & 8 & 20 & 22 \\
      \hline 60 & 10 & 6000 & 8 & 40 & 42 \\
      \hline 160 & 1 & 160 & 18 & 6 & 8 \\
      \hline 160 & 2 & 640 & 18 & 12 & 14 \\
      \hline 160 & 3 & 1440 & 18 & 18 & 20 \\
      \hline 160 & 4 & 2560 & 18 & 24 & 26 \\
      \hline 160 & 5 & 4000 & 18 & 30 & 32 \\
\hline
		\end{tabular}
\qquad
			\begin{tabular}{|c|c|c|c|c|c|}
    \hline $n_h$ & $l$ & $n$ & $k$ & $d(\overline{Z})$ & $d(\overline{X})$  \\
      \hline 360 & 1 & 360 & 38 & 8 & 8 \\ 
      \hline 360 & 2 & 1440 & 38 & 16 & 16 \\ 
      \hline 360 & 3 & 3240 & 38 & 24 & 24 \\ 
      \hline 360 & 4 & 5760 & 38 & 32 & 32 \\ 
      \hline 360 & 5 & 9000 & 38 & 40 & 40 \\ 
      \hline 1800 & 1 & 1800 & 182 & 10 & 10 \\
\hline
		\end{tabular}
	\end{center}
	\caption{Hyperbolic and semi-hyperbolic surface codes based on the $\{4,5\}$-tiling. We give the minimum weights $d(\overline{Z})$ and $d(\overline{X})$ of any logical operator of $X$-type and $Z$-type, the number of qubits $n_h$ of the purely hyperbolic code, the total number of qubits $n$ of the 
(semi)-hyperbolic code, and the parameter $l$ used for the $l\times l$-tiling of every square face.} \label{tab:codes}
\end{table}

\section{Threshold of Hyperbolic Surface Codes with Qubit and Measurement Errors}
\label{sec:num}

We assume that the physical qubits in a code block are subject to a phenomenological $X-Z$ error model. 
This means that prior to each QEC step a qubit undergoes an $X$ error with probability $p$ and a $Z$ error with probability $p$. 
The QEC step itself consists of an instantaneous measurement of all parity checks of the code. 
We will refer to the set of 0 and 1 outcomes of these parity checks as the syndrome. 
We model the noise in this error correction step by assuming that each parity check is obtained perfectly and then independently flipped with some probability $q$. In our numerical studies we restrict ourselves to $q=p$. 

As with the toric code, one repeats the QEC step some $\t$ times and one can infer errors based on this record using a minimum-weight matching algorithm (MWM). The  hyperbolic code lattices are in general not self-dual and the minimal distance of $\overline{X}$ is different from the minimal distance of $\overline{Z}$, see Table \ref{tab:codes}.  Hence one runs the decoder independently on two sets of syndromes, those of $X$-type and those of $Z$-type, to detect and correct for $Z$ and $X$ errors respectively. To correct for $X$ errors we use the description below where $G$ is the graph associated with the dual lattice of the code.

\begin{figure}[htb]
    \begin{center}
        \includegraphics[scale=0.8]{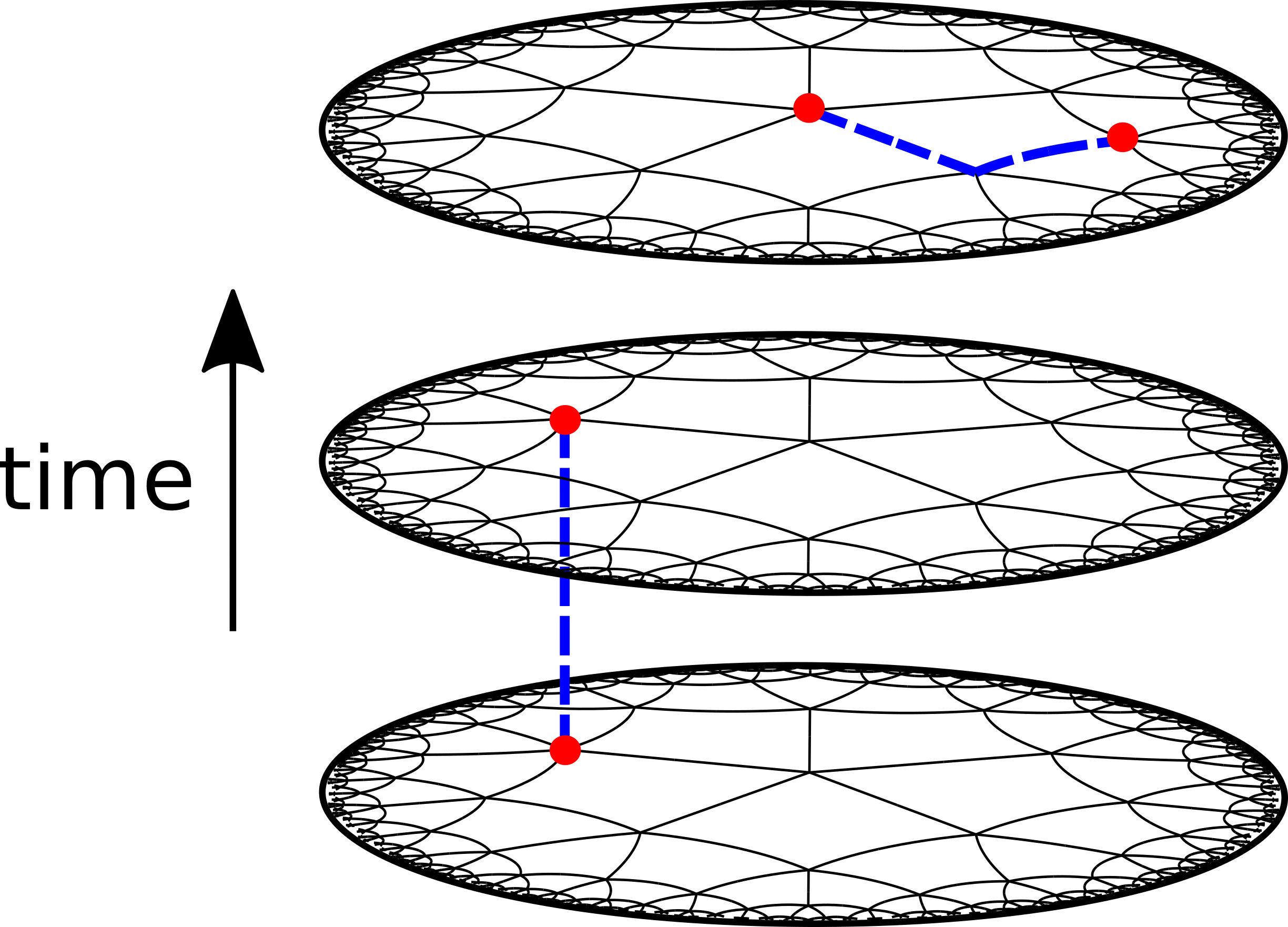}
    \end{center}
    \caption{Minimum weight matching for noisy syndromes in a hyperbolic space. For ease of illustration we are showing the infinite lattice instead of a finite, compactified one. There are three QEC cycles. Marked vertices are indicated by red dots. The result of the MWM is indicated by the blue, dashed lines.}
    \label{fig:noisy_synd_dec}
\end{figure}

When parity check measurements are faulty, one modifies the decoding procedure for noiseless parity checks in a standard manner, first described for the toric code in \cite{dklp}. Let $G=(V,E)$ be the graph associated with the code.
One makes $\t+1$ copies of the graph $G$: each vertex $v$ in copy $G_i$ is connected to the same vertex $v$ in copy $G_{i+1}$ via an edge, obtaining a new graph $G_{\rm time}$, see Figure \ref{fig:noisy_synd_dec}. Each copy represents one QEC cycle in which a qubit error can take place and the entire faulty syndrome is measured.

The decoding algorithm for $Z$ errors then proceeds as follows:
\begin{enumerate}
\item \textbf{Mark vertices:}

    Assume at time $t = 0$ a fictitious round of perfect QEC (no measurement or qubit error and thus all syndromes are 0). 
\begin{itemize}
\item For each QEC cycle at time $1 \leq t \leq T$, mark a vertex when the syndrome is different from the previous time $t-1$.
\item Add a round $t=T+1$ with no syndrome error and mark a vertex when the syndrome is different from the syndrome at $T$.
\end{itemize}

The last round ensures that the total number of marked vertices is even: this step plays the role of ideal decoder and allows one to capture the logical error probability after $T$ rounds of QEC. One can thus perform minimum weight matching on the set of marked vertices:

\item \textbf{Perform MWM on marked vertices:}
For each pair of marked vertices, compute the minimum distance between them using the graph distance in the graph $G_{\rm time}$.
Feed the set of marked vertices along with the minimal distance paths between them to the MWM algorithm.
The algorithm will output pairs of marked vertices such that the sum total of the weight of paths between pairs is minimized.
The inferred error is the shortest path between each pair. This path consists of vertical edges (parity check measurement errors) and horizontal edges (qubit errors),  see Figure \ref{fig:noisy_synd_dec}.
\item \textbf{Deduce residual errors and determine whether a logical error has occurred:}
We infer the errors that remain at time $\t+1$ by projecting the inferred error to the last time step, obtaining a set of only horizontal edges.
A horizontal edge $e$ is an element of this projected set if it was included an odd number of times in the matching. We take the real error that has occurred and project it similarly onto the $T+1$ time-slice. The product of the real and inferred $Z$ error is a closed $Z$-loop and we check whether it is a logical operator by checking whether it anti-commutes with {\em any} of the $\overline{X}_i$ operators. If it anti-commutes, we declare it a logical failure.
\end{enumerate}

For a fixed given $T$ this decoding procedure is applied to stochastically-generated $Z$ errors and repeated $N$ times so that $\overline{P}$ is estimated as $N_{\rm fail} / N$. Next to $\overline{P}$ one can define an effective error probability per QEC round $P_{\rm round}$, where $P_{\rm round}$ is simply defined by the equation $(1-P_{\rm round})^T = 1 - \plog$. The quantity $P_{\rm round}$ can be thought of as the average probability of a logical error occurring at any time step.

When parity checks are noiseless, the decoding routine only has a single round and one marks the syndrome vertices which are 1. Again we output the logical error probability $\overline{P}$ as $N_{\rm fail}/N$. This has already been done for hyperbolic codes in \cite{BT:hyper}. We will present the results for the semi-hyperbolic codes in the following section.

\subsection{Results on Noiseless Parity Checks}

\begin{figure}[htb]
\centering
    \includegraphics[scale=1]{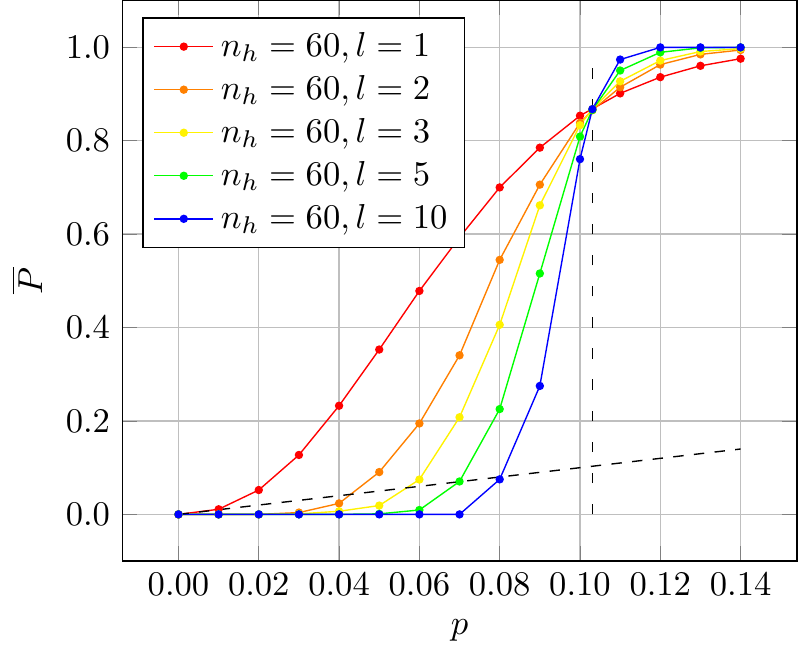}
    \caption{Threshold of a $\{4,5\}$-semi-hyperbolic code family, see Table \ref{tab:codes} with $k = 8$ logical qubits and $l = 1,2,3,5,10$, with noise-free parity checks. 
             The case $l = 1$ is identical to the original hyperbolic code. The vertical, dashed line marks the threshold of the toric code at 
             $10.3\%$ and the diagonal dashed line marks $p = \plog$.}
    \label{fig:semi-hyp_threshold}
\end{figure}

In Figure \ref{fig:semi-hyp_threshold} we present some results for semi-hyperbolic tilings when the underlying hyperbolic tiling is fixed.
The threshold for growing $l$ but fixed $n_h$ tends to be that of the toric code. Intuitively, this result can be understood by observing that this family of codes deviates from a toric code only around a constant number of vertices (see Figure \ref{fig:local_regions}).

\begin{figure}[htb]
\centering
    \includegraphics[scale=1]{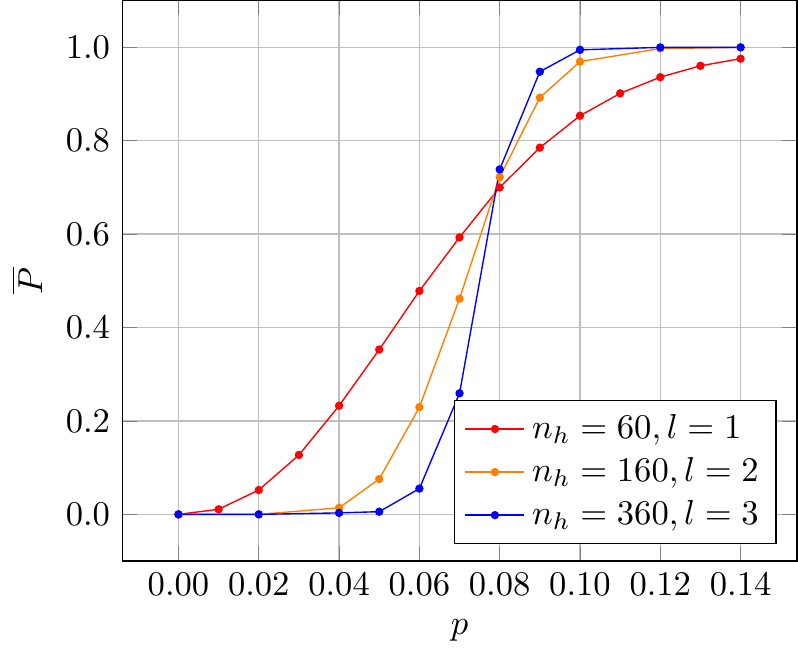}
    \caption{Three $\{4,5\}$-semi-hyperbolic codes $[[60,8,4]]$, $[[640,18,12]]$ and $[[3240,38,24]]$ whose logical error probabilities cross around $7.9\%$. }
    \label{fig:semi-hyp_threshold_diagonal}
\end{figure}

Alternatively, we may define a family of semi-hyperbolic codes where we increase not only $l$ but also the size of the underlying hyperbolic lattice $n_h$.
For example, we can choose $l$ to be proportional to $d_h$. 
This gives a family of codes where the encoding rate $k/n$ is polylogarithmically approaching $0$.
In Figure \ref{fig:semi-hyp_threshold_diagonal} we see that for a code based on a hyperbolic $\{4,5\}$-tiling the lines cross at about $7.9\%$.
This code has a better threshold than  the purely hyperbolic code family \cite{BT:hyper}. We thus see that semi-hyperbolic codes allow for some trade-off between optimizing encoding rate and logical error probability.

\subsubsection{Optimal Value of $T$ for Noisy Parity Check Measurements}

\begin{figure}[htb]
\centering
    \subfloat[Data for $\textrm{[[60,8,4]]}$ ]{
        \includegraphics[scale=0.9]{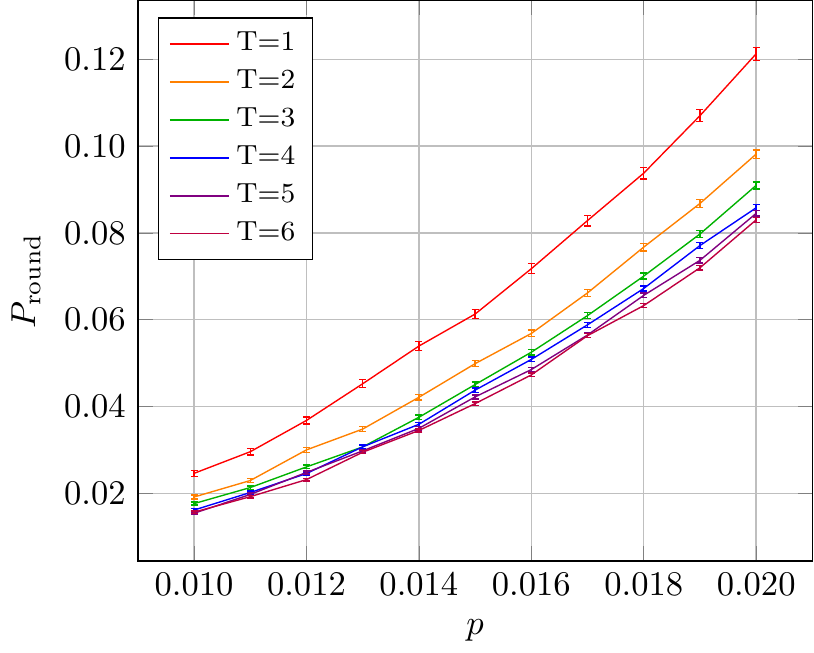}
    }
    \subfloat[Data for $\textrm{[[160,18,6]]}$ ]{
        \includegraphics[scale=0.9]{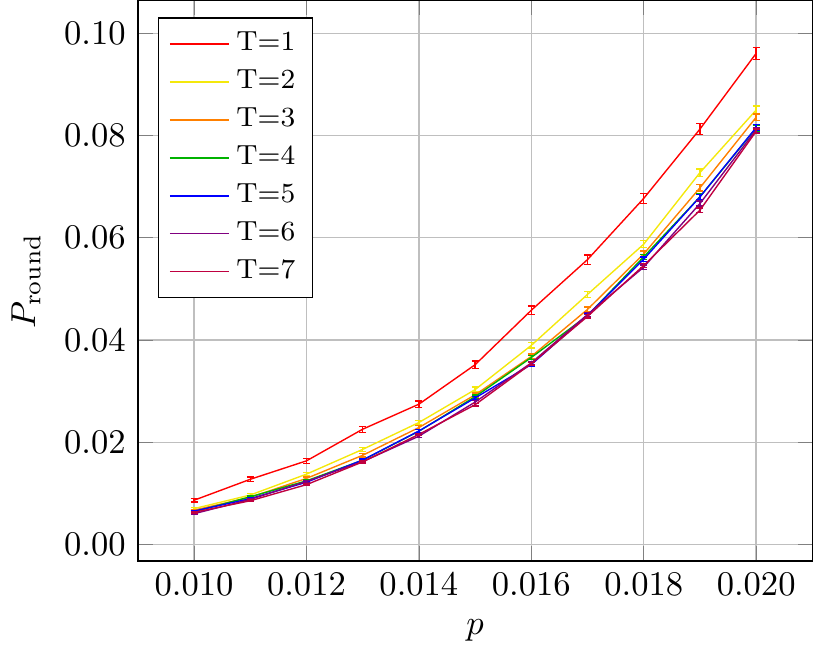}
    }
    \caption{Variation of the logical error probability per round $P_{\rm round}$ with physical error probability for the $\{4,5\}$ lattice.}
\label{fig:time-variation}
\end{figure}

When parity check measurements are noisy, the decoding uses a record of $T$ QEC cycles.
In principle correlating the syndrome record over more rounds of measurements can only improve the efficiency of the decoder per round thus lowering $P_{\rm round}$. %For a model in which parity checks and qubits fail with equal probability one expects that choosing $T \sim d$ for a distance $d$ code. 
In \cite{dklp, toric_code_thresh} it was shown that for an $L\times L$ toric code subject to the previously described noise model, the benefit of taking more than $\t = L$ rounds of syndrome measurement is negligible.
We study the variation of $P_{\rm round}$ with $\t$  in Figure \ref{fig:time-variation}. It can be seen that the improvement between successive rounds steadily decreases: after $T = d$ rounds it becomes rather small. Hence we have used $T=d$ in all further simulations.

\subsection{Cross-Over Behavior and Upper Bounds on $\overline{P}$}
\label{sec:noise}

\begin{figure}
\begin{center}
\includegraphics[scale=1.1]{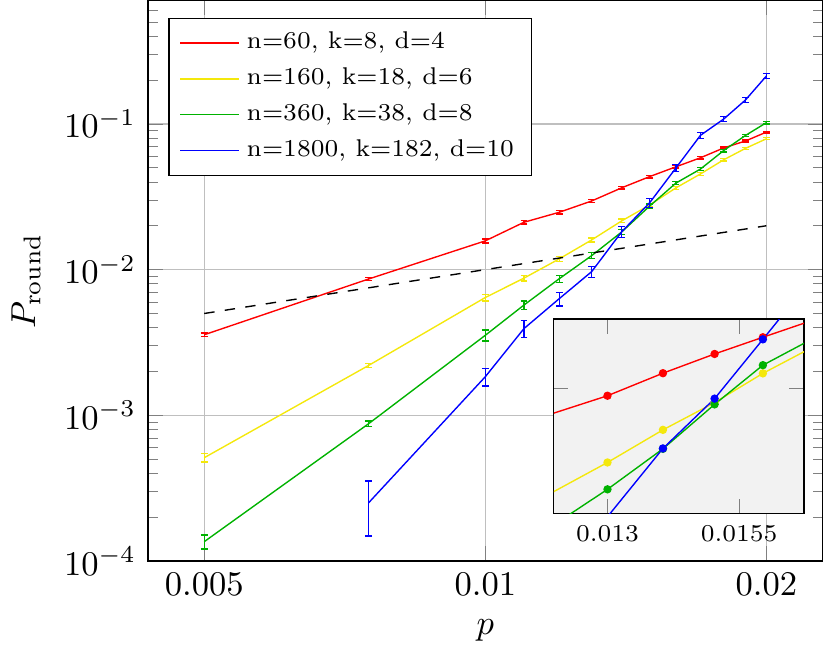}
\end{center}
\caption{
$P_{\rm round}$ vs qubit and measurement error rate $p$. The plot above shows $P_{\rm round}$ for $p$ in the range $0.5\%$ to $2\%$.
The diagonal dashed line marks $P_{\rm round} = p$. The three largest codes seem to cross between $1.3\%$ and $1.55\%$.
}
\label{fig:threshold}
\end{figure}

We study $\overline{P}$ for various hyperbolic codes using the decoding method and the noise model described above. The cross-over point appears to be at around $1.3$\% which can be seen in Figure \ref{fig:threshold}. This cross-over point is somewhat lower than the cross-over point (around $2.5\%$, see \cite{BT:hyper}) for the same codes when parity check measurements are noiseless. This result may be surprising if our intuition is informed by the toric code. If syndromes can be extracted ideally, the threshold of the toric code is at $10.3$\% \cite{toric_code_thresh}.
Changing to the phenomenological error model the threshold drops considerably, to around $3\%$, about a factor of 3 \cite{toric_code_thresh}.

We can try to understand the threshold behavior by calculating an upper bound on the failure probability $\overline{P}$ similar as was done in \cite{dklp}. We assume that the error probability for qubit (horizontal) errors as well as measurement (vertical) errors is $p$. 
We are interested in considering when $\overline{P} \rightarrow 0$ for a growing number of qubits $n$. Let $d$ be the distance of the hyperbolic code. We will use the bound $d \geq c_2 \log n$ for some $c_2$ \cite{BT:hyper}.

Consider error correction for time $T$ for the primal hyperbolic code lattice (hence pertaining to the probability for $\overline{Z}$ errors), generating the space-time graph $G_{\rm time}$ which has $T |V|$ vertices. We consider closed self-avoiding walks in $G_{\rm time}$: closed connected paths which visit each vertex only once. Let such a walk consist of $h$ horizontal edges and $v$ vertical edges. In order for the walk to represent a homologically non-trivial loop it is necessary that the number of horizontal edges is larger than the distance $h \geq d$. Let $\eta_{\rm SAP}^t(h,v)$ denote the number of closed walks (or Self-Avoiding Polygons) with $h$ and $v$ edges with a specific starting vertex at time-slice $t$ in $G_{\rm time}$. Let $E$ be an actual error, a collection of horizontal and vertical edges, and let $E_{\rm min}$ be the chosen correction. Following the arguments in \cite{dklp} one can bound 
\begin{equation*}
\overline{P} \leq |V| \sum_{t=1}^T \sum_v \sum_{h \geq d} \eta_{\rm SAP}^t(h,v)\;\times {\rm Prob}(\mbox{walk(h, v) is contained in }E+ E_{\rm min}),
\end{equation*}
since logical failure (for any of the encoded qubits) can only happen when $E+ E_{\rm min}$ contains a homologically non-trivial loop. We use an upperbound on 
${\rm Prob}(\mbox{walk(h,v) is contained in }E+ E_{\rm min}) \leq 2^{h+v} (p(1-p))^{(h+v)/2}= (4 p(1-p))^{(h+v)/2}$ derived in \cite{dklp} which is independent of the form of the lattice. For noiseless parity checks this upper bound can be used to argue for the existence of a threshold. In this case one has $\overline{P}_{q=0} \leq |V| \sum_{h \geq d}  \eta_{\rm SAP}(h) \alpha^{h}$ with $\alpha=\sqrt{4p(1-p)}< 1$.
One can assume that \cite{barry_wu} $\eta_{\rm SAP}(h) \approx A h^{\gamma-1} \mu_{\rm SAP}^h$ where the connectivity $\mu$, exponent $\gamma$ and the constant $A$ depend on the lattice. For a hyperbolic lattice $\gamma$ is believed to be 1 and bounds exists on $\mu_{\rm SAP}$ \cite{wu_madras}, e.g. for the $\{5,4\}$-tiling it has been proved that for self-avoiding walks we have $2 \leq \mu_{\rm SAW} \leq (36)^{1/3}\approx 3.3$ and $\mu_{\rm SAW} \geq \mu_{\rm SAP}$ (see also \cite{barry_wu}). The fact that the overall prefactor in the upper bound on $\overline{P}$ is expected to be linear in $n$ (i.e. $|V| \sim n$) plays a role in the threshold since the code distance is only logarithmic in $n$.
Using the conjectured form of $\eta_{\rm SAP}(h)$ with $\gamma=1$ and $d \geq \lfloor c_2 \log n \rfloor$ gives 
\begin{equation}
\overline{P}_{q=0} \leq \Theta(n) \sum_{h \geq \lfloor c_2 \log n\rfloor} \left(\alpha \mu_{\rm SAP}\right)^{h}. 
\end{equation}
The upper bound on the r.h.s. is vanishing as $\Theta(n^{\beta})$ for $n \rightarrow \infty$ when 
\begin{equation}
\beta=1+ c_2 \log \left(\alpha \mu_{\rm SAP}\right) < 0.
\label{eq:thresh}
\end{equation}
We cannot estimate the threshold by taking equality in this equation since we do not know of very good bounds on $c_2$ (see \cite{BT:hyper} for a brief discussion). The hyperbolic code threshold for $q=0$ has been found to be lower than the toric code. This might be due to two factors. First of all, $\mu_{\rm SAP}$ is different for a Euclidean versus a hyperbolic lattice, impacting for what $p$ one meets the condition $\alpha \mu_{\rm SAP} < 1$. But possibly more relevant is the fact that for the hyperbolic codes, (\ref{eq:thresh}) tells us that one has to obey the possibly more stringent condition $(\alpha \mu_{\rm SAP})^{c_2} < 1/2$ in order to be below threshold (given that the distance is $\sim \log n$ one needs to overcome to $\sim n$ entropic prefactor). From this perspective one also expects that the phase transition for hyperbolic codes is different from than that of the toric code, rather resembling a Kosterlitz-Thouless transition. 

We can extend these arguments only roughly to the case $q=p$ assuming that in general $ \eta_{\rm SAP}(h,v) \leq {\rm poly}(l) \mu^l$ with $l=h+v$ for some $\mu$. Then assuming that $\alpha \mu < 1$ one can upperbound
\begin{equation}
\overline{P}_{q=p} \leq {\rm poly}(n \log n) (\mu \alpha)^{c_2 \log n}.
\end{equation}
The threshold condition now will depend on the prefactor ${\rm poly}(n \log n)$ which in turn depends on having more precise knowledge about $\eta_{\rm SAP}(h,v)$.

\section{Overhead for a Target Logical Error Probability}
\label{sec:overhead}

A concrete application-oriented goal of using finite-size codes is to determine the qubit overhead given a physical error probability to reach a certain logical error probability for all encoded qubits, see e.g. \cite{CDT:study}.
A simple comparison between (semi)-hyperbolic codes and copies of the toric code can be done by fixing the number of logical qubits $k$, the distance $d$ and compare the number of physical qubits $n$. 
The toric code parameters are $[[2d^2,2,d]]$. To have the same number of encoded qubits we take $k/2$ copies of the toric code, each with distance $d$, so $N_{\rm toric}=kd^2$. For the (semi-) hyperbolic codes that we have studied one has 
$N_{\rm hyper}=\frac{k d^2}{c_2 \log (10 k)}$ assuming the asymptotic rate $k/n\rightarrow 1/10$ and a distance $d=c_2\log n$ for the $\{4,5\}$-hyperbolic code.

In order to get more insight into the possible savings one can numerically compute the maximum error probability $p_{\rm max}(\overline{P}_{\rm target})$ such that $\overline{P} \leq \overline{P}_{\rm target}$ for a surface code. Here $\overline{P}$ is the logical error probablity after $T=d$ QEC rounds when the code distance is $d$.

We have executed this numerical analysis for $\overline{P}_{\rm target}=10^{-5}$, resulting in the values
\begin{eqnarray}
\mbox{[[60, 8, 4]]} \colon \; p_{\rm max}(10^{-5}) \approx1.5\times 10^{-4} \nonumber \\
\mbox{[[160,18,6]]} \colon \; p_{\rm max}(10^{-5}) \approx 9.5 \times 10^{-4}  \\
\mbox{[[360,38,8]]} \colon \; p_{\rm max}(10^{-5}) \approx 1.5 \times 10^{-3} \nonumber 
\end{eqnarray}
In order to compare the performance of the hyperbolic codes with the toric code we will focus on the largest of the three codes which has 38 logical qubits. 
In order to encode 38 logical qubits using the toric code, one needs 19 torii each with $2L^2$ physical qubits. 
If we choose all torii with $L=3$ one has a total of 342 qubits and at $p=1.5 \times 10^{-3}$ numerical data show that 
$\overline{P}= (6.8\pm 0.7) \times 10^{-3}$ after 3 QEC rounds. 
Remember, that $\overline{P}$ is the probability for {\em any} logical qubit to be corrupted.
For $L=4$ one has 608 qubits in total and at $p=1.5 \times 10^{-3}$ numerical runs give the estimate $\overline{P}=(9.3\pm  0.6) \times 10^{-3}$ after 4 QEC rounds. 
Given that all 38 logical qubits encoded in the hyperbolic code with 360 physical qubits have a logical error probability of $10^{-5}$ after $8$ QEC rounds, it clearly outperforms these toric codes.

There is a version of the toric code that we will call the {\em rotated toric code} which has a better scaling between distance and number of physical qubits (it can be obtained by gluing together the boundaries of the rotated surface code).
Taking the set $[0,L]^2 \subset \mathbb{R}^2$ and identifying all points $(x,0)\sim (x,L)$ and $(0,y)\sim (L,y)$  for any $x,y \in [0,L]$ gives a torus.
Instead of tiling it with a square grid in the canonical way we choose the vertices of the tiling to be located at integer points $(x,y)\in \{0,...,L-1\}^2$ for even $x$ and $(x,y-1/2)\in \{0,... ,L-1\}\times \{1/2,... ,L-1/2\}$ for odd $x$.
Edges run diagonally from $(x,y)$ to $(x,y+1/2)$ and to $(x,y-1/2)$.
This procedure gives a square grid on the torus, rotated by 45 degrees.
Note that the shortest non-trivial loop following the edges around the torus has length $L$ while the total number of edges, and hence the number of qubits in the derived code, is $L^2$ as compared to $2L^2$ for the regular toric code.
The number of encoded qubits is still 2 as there are two independent, non-trivial loops.

Using 19 rotated toric codes we can either use $L=4$ or $L=6$ amounting to 304 and 684 physical qubits resp. For $L=4$ the logical error probability at $p=1.5 \times 10^{-3}$ is numerically estimated to be $\overline{P}= (2.3\pm 0.1) \times 10^{-2}$ after 4 QEC rounds.
For $L=6$ the logical error probability at $p=1.5 \times 10^{-3}$ is numerically estimated to be $\overline{P}= (7.0\pm 0.2) \times 10^{-4}$ after 6 QEC rounds.

In order to further estimate the scaling of the logical error probability we write down an approximate model for the logical error probability in Section \ref{sub:approximation} which we use in Section \ref{sub:overhead} as the basis for further comparison.

\subsection{Approximation for $\plog$ in the low error probability limit}\label{sub:approximation}

We focus on getting an expression for the logical error probability $\plog$ when the physical error probability $p$ is low compared to the noise threshold, assuming a minimum-weight decoding method. This approach has been used for the surface code in \cite{fowler}. We first consider the case of noiseless parity checks.

The logical error probability for, say, a $\overline{Z}$ error is given by summing the probabilities of any $Z$-error to occur, times the probability of the decoder to fail on this error. In order for a minimum-weight decoder to fail, the weight of the error $E$ must be at least $|E| \geq \lceil d/2 \rceil$. In other words:
\begin{equation*}
    \plog = \sum_{E:\; |E| \geq  \lceil d/2 \rceil} P(\mbox{MWM fails on } E)\; p^{|E|} (1-p)^{n-|E|}.
  \end{equation*}
We are interested in the small $p$ regime of $\plog$ which is a polynomial in $p$. We will thus retain only the lowest order $p^{\lceil d/2 \rceil}$ term, i.e. $\overline{P} \approx  P_0^{q=0}$ where $P_0^{q=0} \equiv \sum_{E:\;|E|\geq \lceil d/2 \rceil} P(\mbox{MWM fails on } E)\; p^{|E|}$.
The errors of weight $\lceil d/2 \rceil$ on which the minimum-weight decoder fails are exactly those where all the support of the error is in the support of a weight-$d$ logical operator. There are $d\choose d/2$ of such errors.
If $d$ is odd then the MWM-decoder will fail with probability 1. If $d$ is even, then there are two decodings that either lead to a successful decoding or a failure.

Assuming that the decoder will choose randomly among these, the probability of failure is $1/2$ in this situation. Since the logical operators in the hyperbolic surface code will overlap on qubits, we can only upperbound $P_0^{q=0}$ as
\begin{equation}\label{eqn:approx_P_log}
    P_0^{q=0} \leq N_d \left( \frac{3}{4} - \frac{1}{4} (-1)^{d} \right)
    {d\choose \lceil d/2 \rceil}  \; p^{\lceil d/2 \rceil }
\end{equation}
where $N_d$ is the number of logical operators of weight $d$. The right hand side of (\ref{eqn:approx_P_log}) can be used to approximate the error probability when syndrome measurements are ideal.

\begin{figure}[htb]
    \centering
    \subfloat[Toric code with $L=6$.]{
        \includegraphics[scale=0.9]{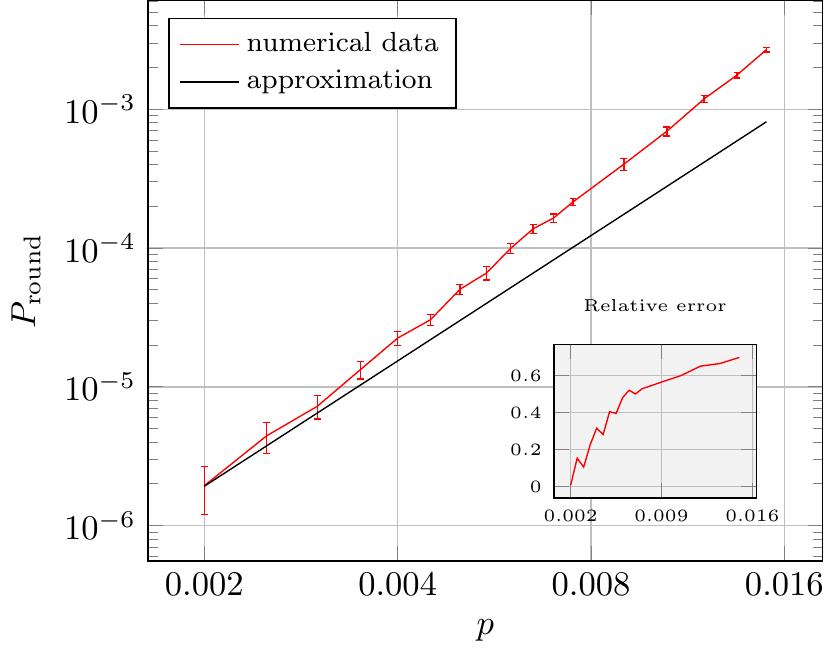}
    }
    \subfloat[$\{4,5\}$-hyperbolic  code with $n=160$.]{
        \includegraphics[scale=0.9]{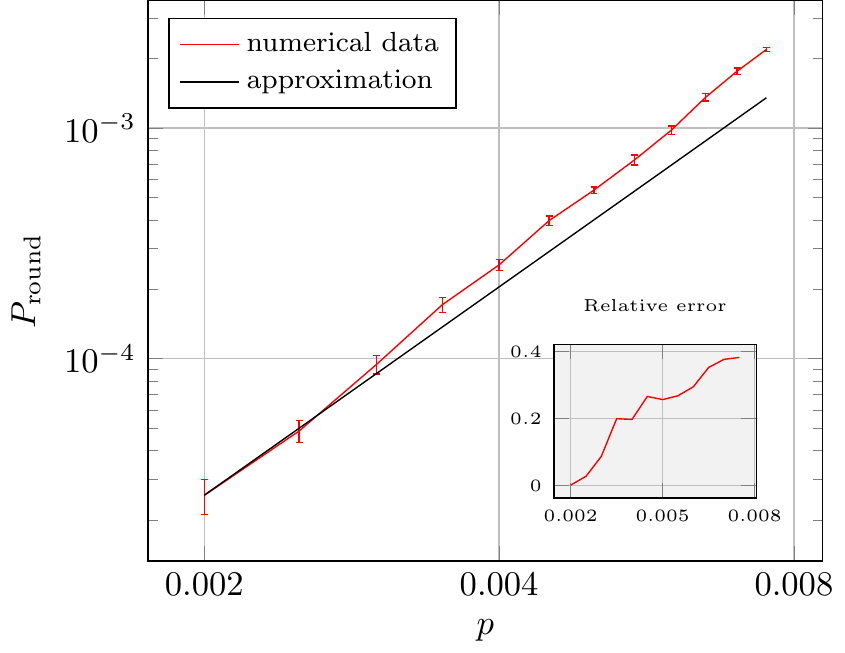}
    }\\
    \caption{Comparing numerical estimates for $P_{\rm round}$ (red) with the heuristic approximation in (\ref{eqn:approx_P_log_noisy}) (black). The relative error is the absolute difference between the numerical value and the approximation divided by the numerical value.}
\label{fig:approximation_formula}
\end{figure}

For noisy parity check measurements, taking again the low $p$ limit, we can apply the same reasoning and only consider the lowest weight error configurations that can possibly lead to a logical failure.
These lowest weight errors must then lie within a single time slice, so that one has
\begin{equation}\label{eqn:approx_P_log_noisy}
    P_0^{q=p} \leq T N_d \left( \frac{3}{4} - \frac{1}{4} (-1)^{d} \right)
    {d\choose \lceil d/2 \rceil}  \; p^{\lceil d/2 \rceil }.
\end{equation}
For the hyperbolic codes there is no  formula for $N_d$.
However, one can compute $N_d$ efficiently, see \ref{sec:dist_comput} and the results for various codes. The approximation in (\ref{eqn:approx_P_log_noisy}) agrees well with data obtained from numerical simulations. This can be seen in Figure \ref{fig:approximation_formula} where we compute the per-round 
approximation in (\ref{eqn:approx_P_log_noisy}) versus the numerical per-round logical error probability $P_{\rm round}$.

\begin{figure}[htb]
\centering
    \includegraphics[scale=1]{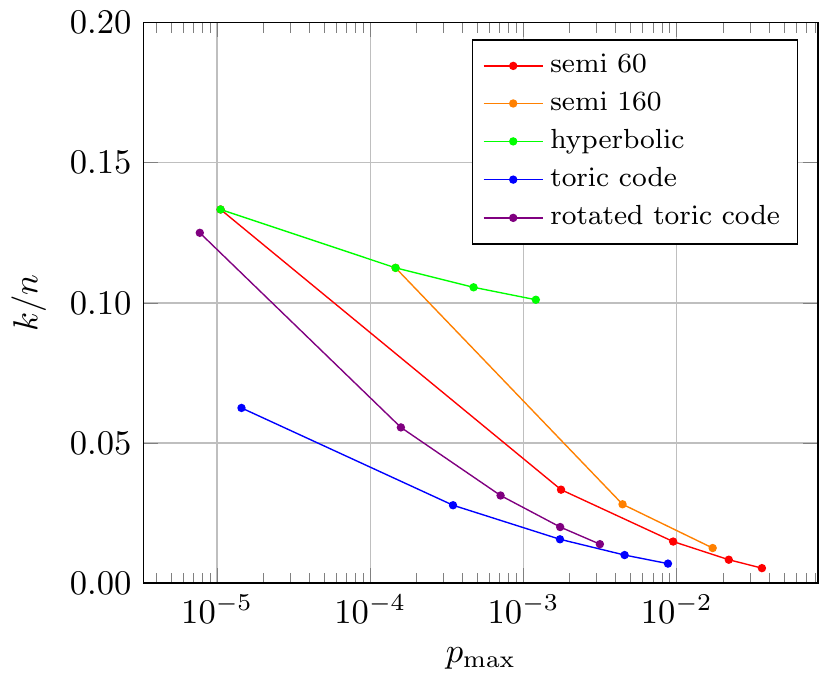}
    \caption{Overhead for different code families. 
        The value of $p_{\max}(10^{-8})$ for various codes.
        The semi-hyperbolic codes in this figure are derived from a $\{4,5\}$-lattice with $n = 60$ and $n= 160$ and $l=2, 3, 4$ etc. 
        The hyperbolic codes are derived from a $\{4,5\}$ lattice with $n = 60,160,360,1800$.
        The toric codes considered here have distance $L=4,6,8,10,12$.}
    \label{fig:overhead}
\end{figure}

\subsection{Overhead of Semi-Hyperbolic Surface Codes}\label{sub:overhead}
Equation \ref{eqn:approx_P_log} can be used to analyze the semi-hyperbolic code family in the regime where the physical error rate $p$ is low.
To compare the overhead in physical qubits we fix $P_{\rm round}$ and determine the maximum physical error probability $p_{\max}(P_{\rm round})$. 
This value for $P_{\rm round}$ was chosen such that the corresponding $p_{\max}$ is in a regime where the approximation formula is valid for all lattices considered here.
In Figure \ref{fig:overhead} we plot the encoding rate $k/n$ against $p_{\max}$ for different code families with $P_{\rm round}=10^{-8}$. 
We ran Monte Carlo simulations for higher values of $p$ to ensure that the approximation did not deviate by more than $10\%$ from the numerical value.
Once this is established, we assume that the approximation will only become better with lower $p$.

For a fixed encoding rate, we see that the semi-hyperbolic codes can offer better protection against errors than the toric code.
For example, in Figure \ref{fig:overhead} we see that for  $p_{\max} = 1.7\times 10^{-3}$ we can choose between copies of the $L=8$ toric code with $k/n = 0.0156$, the rotated toric code with $L=10$ and $k/n = 0.02$, the semi-hyperbolic code with $l=2$ from a $\{4,5\}$ lattice with 60 edges with $k/n = 0.03$ or a hyperbolic code with $k/n = 0.1$.

\section{Retrieving and Writing Qubits to Storage and Dehn Twists for Hyperbolic Surface Codes}
\label{sec:logic}

In this section we present some schemes to realize logical operations on the encoded qubits without increasing connectivity between qubits or losing the overhead advantage of the (semi)-hyperbolic codes. Since we are not aware of a way to fault-tolerantly implement a universal gate set, or even the full Clifford group via, say, code deformation within (semi)-hyperbolic codes, we envision an architecture similar to that of a classical processor. 
At every instant, only a subset of the qubits is undergoing computation while the rest are in storage. The storage medium here is a (semi)-hyperbolic code while the computational space is thought of as a few blocks of 2D surface or color codes with magic state distillation capabilities. 

\subsection{Lattice Code Surgery}

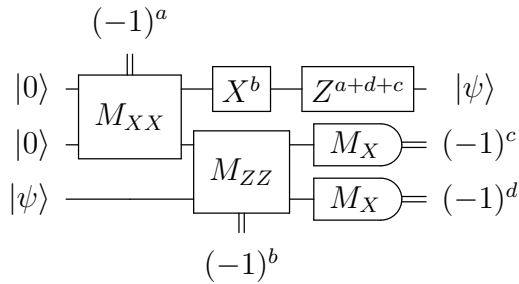
\begin{figure}[hbt]
\centering
\vspace{1em}
\mbox{
\Qcircuit @C=.4em @R=.4em{
\lstick{} & \ustick{(-1)^a} \\
\lstick{\ket{0}} & \multigate{1}{M_{XX}} \cwx &  \gate{X^{b}} &\gate{Z^{{a+d+c}}}& \qw & \rstick{\ket{\psi}}\\ 
\lstick{\ket{0}} & \ghost{M_{XX}}  & \multigate{1}{M_{ZZ}}  & \measureD{M_X} & \rstick{(-1)^c} \cw\\
\lstick{\ket{\psi}} &\qw & \ghost{M_{ZZ}} & \measureD{M_X} & \rstick{(-1)^d} \cw \\
\lstick{} & & \dstick{(-1)^b} \cwx } }\vspace{1em}
\caption{One possible circuit to realize one-bit teleportation via measurements. It uses one ancillary qubit and two weight-two joint measurements. The boxes containing $M_{XX}$ and $M_{ZZ}$ indicate a joint measurement of the two qubits involved.}
\label{fig:onebit_tele}
\end{figure}

In order to read or write qubits to storage, one requires the following operations on individual qubits without affecting the protection of other logical qubits:
\begin{itemize}
\item Measure qubit in storage in $Z$ or $X$ basis (and thus also reset individual qubit). This step can be accomplished by performing a joint $ZZ$ (resp. $XX$) measurement on the stored qubit and a qubit in the computational space which is initialized to $\ket{0}$ (resp. $\ket{+}$).
\item Retrieve qubit from storage into computational space or write a qubit to storage. In order to extract logical qubits from storage to the computational space, one can implement one of the standard one-bit teleportation circuits using again $XX$ and $ZZ$ measurements (see \autoref{fig:onebit_tele}).
\end{itemize}

The requirement to perform logical $XX$ or $ZZ$ measurements between a stored qubit and a qubit in the computational space can be fullfilled using the technique of lattice code surgery which is usually applied between sheets of surface code \cite{horsman}. It can be straightforwardly generalized to closed surfaces which encode more logical qubits as we shall now argue.

\begin{figure}[htb]
    \centering
    \subfloat[ ]{
        \includegraphics[width=.3\textwidth]{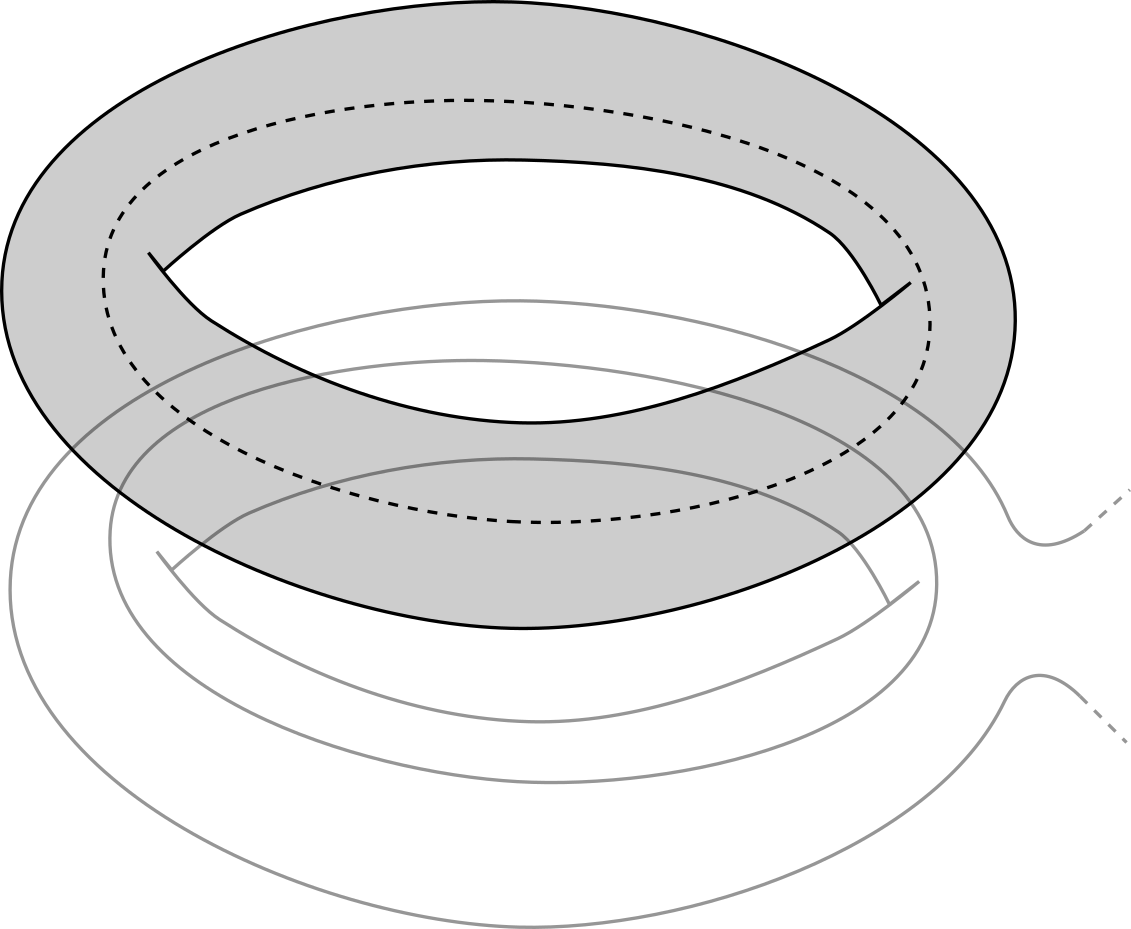}
    }\hspace{2cm}
    \subfloat[]{
        \includegraphics[width=.25\textwidth]{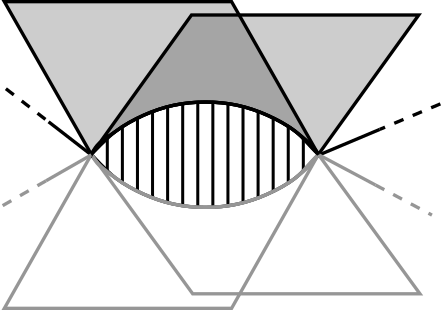}
    }
\caption{(a) Positioning of the ancillary toric code (grey, on top) with respect to the storage transfer zone (white, on bottom) with $Z$ logical operators facing each other to realize a $ZZ$ measurement. (b) Local configuration of the merged lattices after measuring qubits in the support of the logical $Z$ operator in facing pairs. The paired qubits lie on the two curved edges and between them is a 2-edge face (striped) glued perpendicular to both surfaces. Note that the merger leads to $X$-checks of weight-8 (by adding a layer of qubits in between the torii one can reduce this to weight 5).}
\label{fig:surgerytori}
\end{figure}

Instead of matching boundaries to one another, one has to match logical operators in the middle of the surfaces but this does not affect the overall procedure.
Figure \ref{fig:surgerytori} (a) represents a possible configuration for a logical $ZZ$ measurement.
Two handles are on top of each other with two matching logical $Z$ operators facing each other.
One would then measure pairs of facing qubits in the support of these two operators in the $Z$ basis.
Their product gives the outcome of the joint $ZZ$ measurement and the two handles are merged. The new measurements of pairs of qubits does not commute with the local $X$-checks of the separate surfaces, so they get replaced by products of these $X$-checks which do commute \footnote{One can view this as a standard application of the Knill-Gottesman theorem by which one keeps track of the stabilizer group after performing Pauli measurements.}. The result is a merged lattice which is no longer the discretization of a 2-manifold as it contains edges adjacent to three faces, see Figure \ref{fig:surgerytori} (b).
One can observe that the two logical $Z$ operators become equivalent under the application of the new two-edged faces (which are elements of the new stabilizer group). Furthermore, the two corresponding logical $X$ operators have to be merged in order to commute with those two-edged faces.
Error correction can be carried out in this merged phase by using the previous unchanged $Z$-checks for $X$ errors and the new merged $X$-checks for $Z$ errors. Once the outcome is known, one splits back the two handles by measuring the previous, separated $X$-checks.

This results in correlated $Z$ errors that have to be corrected in a correlated fashion, the same way as in the standard surface code lattice surgery.
It is important to note that the fact that the code contains some other encoded qubits which can have a representative logical operator supported on the modified region is not a concern.

The correction, restricted to the storage space consists of applying Pauli $Z$ operators to a subset of the qubits forming the measured $Z$ logical operator, $Z_{\rm meas}$.
Denote $S_{\rm corr}\subset{\rm Supp}(Z_{\rm meas})$ this subset.
Equivalently, the complement of this subset, $S_{\rm corr}^\prime = {\rm Supp}(Z_{\rm meas})\setminus S_{\rm corr}$ can be used for the correction.
Take some logical operator of another logical qubit of the code, $X_{\rm other}$.
Since $X_{\rm other}$ commutes with $Z_{\rm meas}$, it has to overlap with ${\rm Supp}(Z_{\rm meas})$ on an even number of qubits.
That implies that the two choices of correction, $S_{\rm corr}$ or $S_{\rm corr}^\prime$, both have the same effect on it.
Both either flip its sign or both leave it invariant.
Moreover, using the $X$-checks lying on ${\rm Supp}(Z_{\rm meas})$, one can move around the loop where $X_{\rm other}$ intersects $Z_{\rm meas}$.
So there is another representative for $X_{\rm other}$ that is unmodified by the correction.
The correction just enforces that all other representative are equivalent to an unmodified one.
 
\subsection{Movement of Qubits in Storage: Dehn Twists}

Needing ancillary qubits and connecting these to the storage qubits is a concern for the overall connectivity and overhead of our proposal. Let us introduce two measures of connectivity. There is an {\em instantanteous qubit degree} which is the number of other qubits that a qubit has to interact with (for doing parity check measurements) at a certain point in time. We would like this degree to be a small constant throughout our schemes. 
Besides this notion there is a {\em cumulative qubit degree} which measures the total number of different qubits that a qubit has to interact with over time. For hardware with fixed connections this cumulative qubit degree should ideally be a small constant as well. For hardware which allows for switching (e.g. switches in a photonic network) the cumulative qubit degree could be allowed to grow. 

If we were to decide to only have one (logical) ancillary qubit linking every storage qubit to the computational qubits then we will blow up the cumulative qubit degree of this ancillary qubit (cumulative degree scaling with the number of logical qubits $k$). On the other hand if we use one ancillary qubit for each storage qubit we give up the overhead advantage given by the (semi)-hyperbolic code. This is why we need a technique to move qubits around in storage, allowing us to read or write qubits from storage only at certain locations. Such storage medium will not be a random access memory since the retrieval of encoded qubits depends on where they are stored in the memory. 
For the movement technique we propose the use of Dehn twists which is a code deformation technique using the topological nature of the code to implement operations \cite{Koenig:Dehntwists, book:mapping}.
In a nutshell, Dehn twists allow us to perform CNOTs between the two qubits of one handle as well as exchanging pairs of qubits between handles.
This then allows us to have designated zones for transfer from and to the computation space and move storage qubits to these zones when needed.

Our movement proposal in the form of Dehn twists leads to a growing cumulative qubit degree of some of the physical qubits in the code (cumulative degree scaling with distance $d\sim \log n \sim \log k$).
 In Section \ref{sec:reduct} we suggest a way in which one can modify this method leaving the cumulative degree of qubits constant at the expense of using additional space (qubits).

\begin{figure}[htb]
\centering
\includegraphics[width=.3\textwidth]{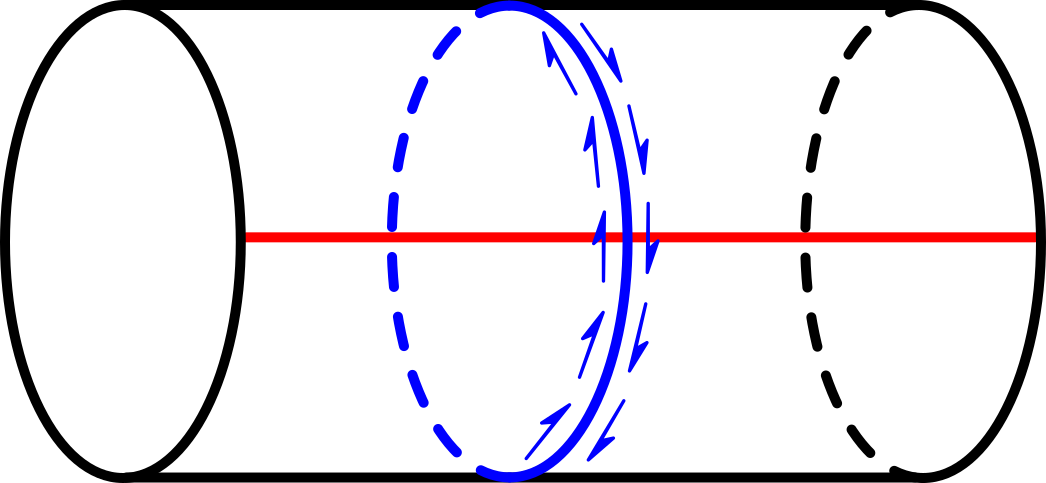}
\raisebox{2.3em}{\quad$\longrightarrow$\quad}
\includegraphics[width=.3\textwidth]{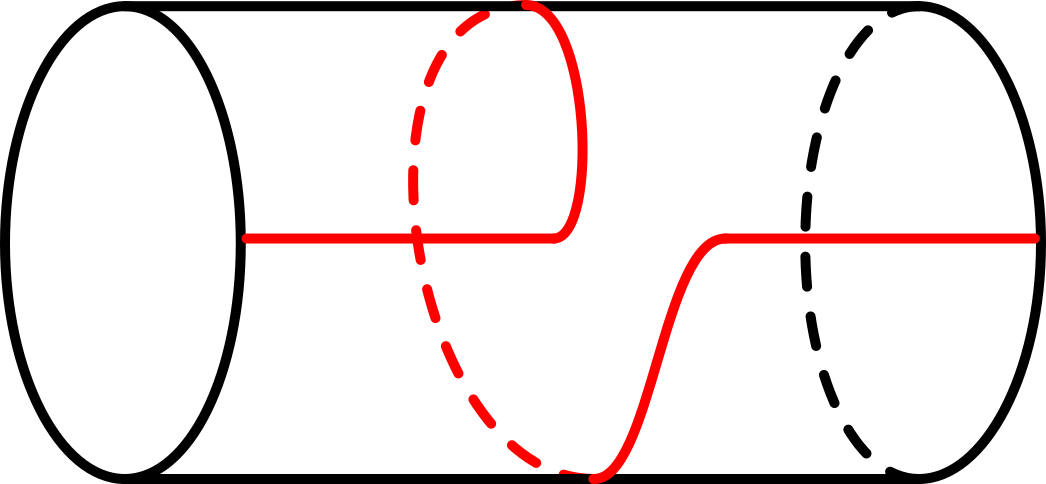}
\caption{The action of a Dehn twist along the arrowed (blue) loop. It adds this loop to the (red) path crossing it.}
\label{fig:dehntwistdeform}
\end{figure}

\begin{figure}[hbt]
\centering
\includegraphics[width=.7\textwidth]{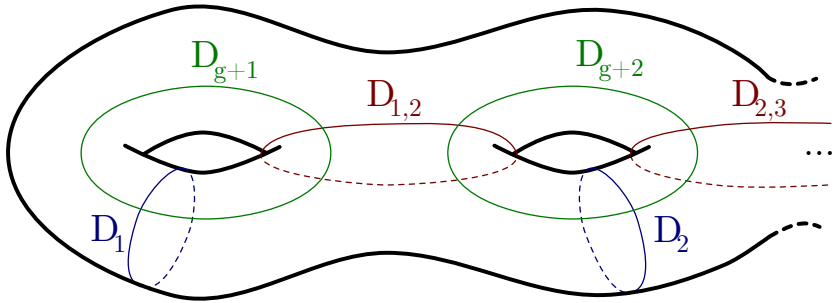}
\caption{A generating set of loops for Dehn twists on a surface with $g$ handles. Each handle hosts two qubits, and at the $k$th handle we label the qubits $q_{2k-1}$ and $q_{2k}$. We choose the convention that $\overline{X}_{q_{2k-1}}$ is supported on the loop (on the dual lattice) labelled $k$ and so $\overline{X}_{q_{2k}}$ is supported on the loop $k+g$ (on the dual lattice). That implies that $\overline{Z}_{q_{2k-1}}$ is supported on the loop $k+g$ and $\overline{Z}_{q_{2k}}$ on the loop $k$.}
\label{fig:dehntwistloops}
\end{figure}

A Dehn twist is a homeomorphic deformation of a surface, that is considered here to be compact and having $g$ handles (genus $g$).  Dehn twists on a closed surface $S$ are known to generate the full mapping class group ${\rm MSG}(S)$ of the surface \cite{book:mapping}.
The idea is to twist the surface along a non-contractible loop as shown in \autoref{fig:dehntwistdeform}. This has the effect of adding this non-contractible loop to any other loop that crosses it.

For a surface $S$ we are interested in the effect of Dehn twists on the first homology group $H_1(S,\mathbb{Z}_2)$ as elements in  $H_1(S,\mathbb{Z}_2)$
correspond to the logical $Z$ operators in our code. In other words, we only count how many times a loop wraps around a handle modulo two, so we identify loops with vectors in $\mathbb{Z}_2^{2g}$ using the bases of $2g$ loops shown in \autoref{fig:dehntwistloops} ($D_1$ to $D_{2g}$).
This space can be equipped with a standard symplectic form
counting the number of crossings modulo two between loops. Acting on this space Dehn twists generate the symplectic group $\textrm{Sp}(2g,\mathbb{Z}_2)$ as they preserve the number of crossing modulo two between loops. A possible generating set of size $3g-1$ for the full group is given in \autoref{fig:dehntwistloops}.

\begin{figure}[htb]
\centering
\includegraphics[width=.15\textwidth]{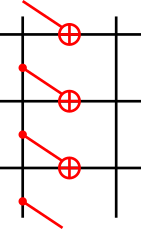}\raisebox{5em}{\quad$\longrightarrow$\quad}
\includegraphics[width=.15\textwidth]{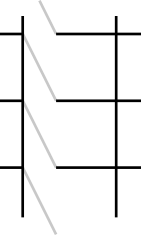}\raisebox{5em}{\quad$\longrightarrow$\quad}
\includegraphics[width=.15\textwidth]{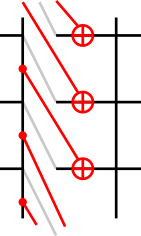}\raisebox{5em}{\quad$\longrightarrow\quad\cdots$}
\caption{The first two steps of a Dehn twist on a square lattice toric code. The qubits are placed on the vertical and horizontal edges, each face is a $Z$-check and each vertex is a $X$-check. The subsequent steps are similar but take into account that the middle row of qubits is gradually displaced ``downwards''.}
\label{fig:dehntwistcnots}
\end{figure}
\begin{figure}[htb]
\centering
\includegraphics[width=.12\textwidth]{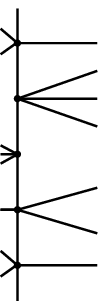}\raisebox{6.5em}{\quad$\longrightarrow$\quad}
\includegraphics[width=.12\textwidth]{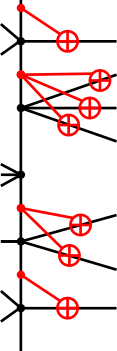}\raisebox{6.5em}{\quad$\longrightarrow$\quad}
\includegraphics[width=.12\textwidth]{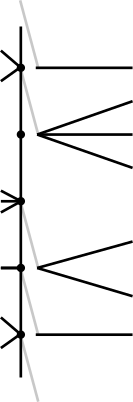}\raisebox{6.5em}{\quad$\longrightarrow\quad\cdots$}
\caption{For the $\{4,5\}$-tiling the Dehn twist procedure has to be slightly generalized. One chooses a non-trivial $\overline{Z}$-loop. The edges sticking out to one side of this loop form the support for $\overline{X}$ of the other qubit of the handle. Instead of having always exactly one edge sticking out to the right (see \autoref{fig:dehntwistcnots}), there can now be between zero and three edges. The modification is then to just adapt the number of target qubits for the CNOTs to this number. At intermediate steps of the Dehn twist one can observe that the $X$-checks have weight varying between 2 and 8.}
\label{fig:dehntwistsemiH}
\end{figure}

It turns out that this kind of continuous deformation has a direct analog for the tiled surfaces of homological codes, so in particular (semi)-hyperbolic codes. A simple example to explain the procedure is that of the toric code.
Using $d$ parallel CNOTs it is possible to ``dislocate'' the lattice by one unit along a loop as shown in \autoref{fig:dehntwistcnots}.
Repeating the step $d$ times with CNOTs which stretch between qubits over a longer and longer range, will bring back the lattice to its initial configuration.
Tracking what happens to a $Z$ or $X$ logical operator which crosses this loop, one can easily see that the procedure acts as a CNOT on the logical operators.
The control qubit $\overline{X}_{\rm control}$ intersects the loop around which the Dehn twist is done on one vertical qubit.
The successive steps apply CNOTs with this qubit as control and qubits of the $\overline{X}_{\rm target}$ parallel to the loop as target.
This gradually propagates $\overline{X}_{\rm control}$ to
$\overline{X}_{\rm target}$.
Symmetrically, $\overline{Z}_{\rm target}$ intersects the loop on one horizontal qubit and the CNOTs propagate it to $\overline{Z}_{\rm control}$ running around the loop.

For our $\{4,5\}$ lattice a slight generalization of this circuit has to be done and is shown in \autoref{fig:dehntwistsemiH}.

\begin{figure}[htb]
\centering

\raisebox{-3em}{$D_k =$}\qquad\quad \raisebox{-2em}{
\Qcircuit @C=1em @R=1em{
\lstick{q_{2k-1}} & \targ &  \qw\\
\lstick{q_{2k}} & \ctrl{-1} & \qw
}}\quad\quad \raisebox{-3em}{$D_{k+n} =$} \qquad\quad
\raisebox{-2em}{\Qcircuit @C=1em @R=1em{
\lstick{q_{2k-1}} & \ctrl{1} &  \qw\\
\lstick{q_{2k}} & \targ & \qw
}}\quad\quad \raisebox{-3em}{$D_{k,k+1} =$} \qquad\quad
\mbox{\Qcircuit @C=1em @R=1em{
\lstick{q_{2k-1}} & \targ & \targ &\qw&\qw\\
\lstick{q_{2k}} & \ctrl{-1} & \qw & \ctrl{1}&\qw \\
\lstick{q_{2k+1}} & \targ & \qw & \targ &\qw\\
\lstick{q_{2k+2}} & \ctrl{-1} & \ctrl{-3} &\qw & \qw
}}

\caption{The circuits realized by the three type of generators for the Dehn twist transformations. The labelling of the Dehn twists and the qubits is the one detailed in \autoref{fig:dehntwistloops}.}
\label{fig:dehntwistcircuits}
\end{figure}
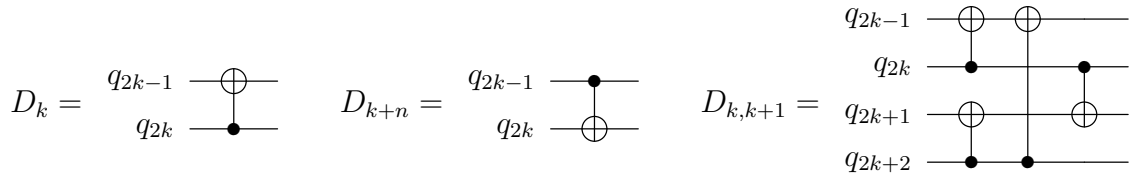

The question is then what useful operations on the logical qubit space these Dehn twists give us access to.
It is easy to verify the action of the generating set of Dehn twists, see \autoref{fig:dehntwistcircuits}.
For our purpose, we can see that nine Dehn twists can be used to swap the qubits of two handles using the circuits in Figure \ref{fig:dehntwistcircuits} to construct SWAP operations from 3 CNOTs. By considering a larger generating set the number of Dehn twists can be reduced to seven. This can be checked by a computer.

\subsubsection{Reducing Connectivity}
\label{sec:reduct}

In each step of the Dehn twist  the (instantaneous) qubit degree is $O(1)$. The cumulative degree of the qubits on the loop along which 
one does a Dehn twist is $O(d)$, with $d$ being the length of this loop.
In the case of hyperbolic codes, this is logarithmic in the total number of physical qubits which is an improvement over losing all overhead or having 
cumulative qubit degree scaling with $k$ by employing read/write ancilla qubits.

The temporal overhead of a Dehn twist, if one applies one round of error correction ($O(d)$ steps in time) between each step is $O(d^2)$.
The cumulative qubit degree can be reduced using the following variation.

We can use an extended region to spread the effect of the twist and lower the connectivity requirements as well as the temporal overhead.
\begin{figure}[htb]
\centering
\includegraphics[height=.23\textwidth]{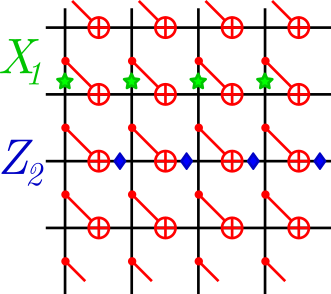}\raisebox{4em}{\quad$\longrightarrow$\quad}
\includegraphics[height=.23\textwidth]{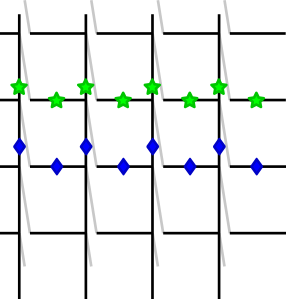}\raisebox{4em}{\quad$\Longleftrightarrow$\quad}
\includegraphics[height=.23\textwidth]{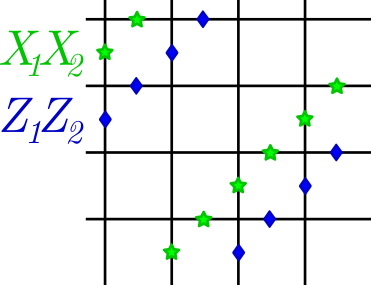}
\caption{Extended Dehn twist on a distance 4 toric code. One does the 4 Dehn twist steps in parallel on $d$ parallel rows of the lattice. Green stars indicate the logical $X_1$ operator and how it transforms to $X_1X_2$. Blue lozenges indicate the logical $Z_2$ operator and how it transforms to $Z_1Z_2$.}
\label{fig:dehntwistextended}
\end{figure}
\begin{figure}[htb]
\centering
\includegraphics[width=.2\textwidth]{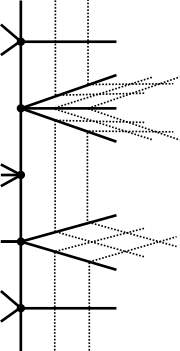}
\caption{Given some initial loop in the $\{4,5\}$-lattice, it is possible to add parallel loops to it. The dotted lines are added qubit edges that make a more fine-grained lattice in the direction mostly ``perpendicular'' to the original loop. This can be somewhat problematic when the original loop takes ``sharp'' turns as in the middle of this example (where there is no qubit edge sticking out to the right). In this face one potentially adds a way for a $\overline{Z}$ to cut a corner and that might decrease the distance by one. One should verify such properties in specific examples of interest.}
\label{fig:extendedsemiH}
\end{figure}
As shown in \autoref{fig:dehntwistextended}, one can choose $d$ parallel loops, and apply in parallel one step of the Dehn twist on each of the loops.
This effectively realizes a Dehn twist in one go in an extended region.
The connectivity required for this extended Dehn twist is constant and doing one round of error correction after this gives a total temporal overhead of $O(d)$.

The only concern is how to adapt this to the $\{4,5\}$-lattice.
There is no a priori guarantee that it is possible to find $d$ parallel loops.
But one can make use of semi-hyperbolic modifications to help with this, basically creating more space for the twist region.
Starting from a chosen $\overline{Z}$-loop, one can add parallel loops by adding qubits in the faces to one side of the loop as shown in \autoref{fig:extendedsemiH}.
This does not completely guarantee that one will create enough parallel
columns as the lattice is expanding to the right.
Because of this, the parallel loops will grow in size demanding more
twisting to complete the full operation.
That said, if one step on the extended region is not enough one can repeat the step on the extended region. So if one step on the extended region twists the lattice for a fraction of the distance, then only a constant number of steps will be needed and the cumulative degree of qubits will remain constant. Also, the total time overhead will be $O(d)$.

\section{Discussion}
\label{sec:discuss}

\begin{figure}[htb]
\centering
\includegraphics[width=.45\textwidth]{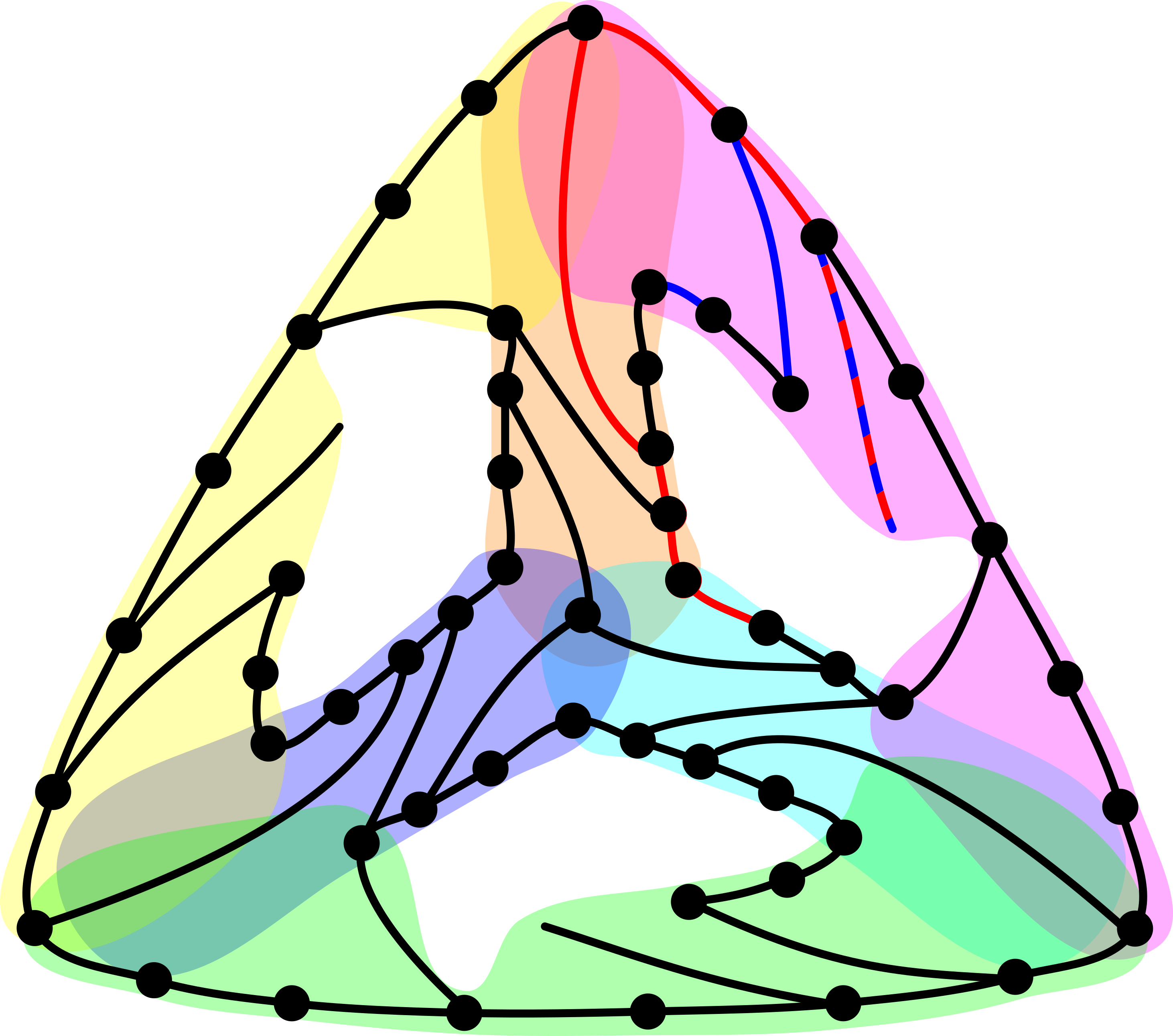}
\qquad
\includegraphics[width=.45\textwidth]{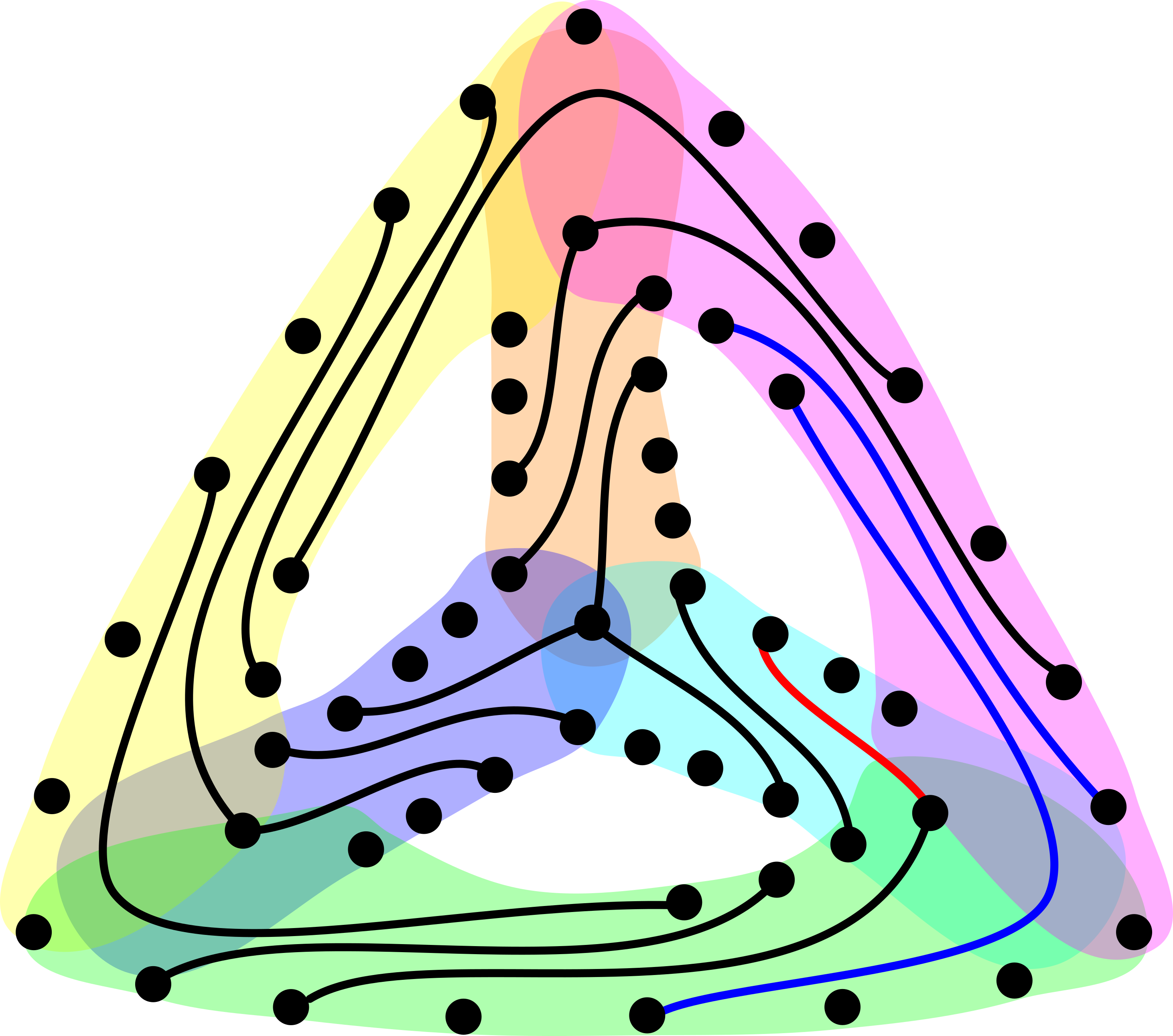}
\caption{(Color online) The Klein quartic, mentioned in Section \ref{sec:HSC}, sliced through so one distributes the qubits in two planar layers shown left and right. We choose the $\{3,7\}$ representation with 84 qubits on the edges and weight-3 $X$-checks (56 of them) and weight-7 heptagonal $Z$-checks (24 of them). The surface has genus 3. Most of the vertices are on the boundary of the layers, and so are represented in both layers even if they only join edges from one layer. They help in visualizing heptagons on the right layer. Few vertices (five) are sitting in the middle of one or the other layer and so are only represented on their layer. The three edges that are dangling in the left layer, are linked to the three degree-2 vertices in the right layer. Red edges indicate one $\overline{Z}$ and blue edges the corresponding $\overline{X}$ which overlap on single bi-colored edge. All vertices (and $X$-checks) are represented in the figure but the figure leaves some ambiguity about those heptagons which act on qubits in both layers.}
\label{fig:KQbilayer}
\end{figure}

Given that the topology of hyperbolic surface codes is that of a surface with many handles, it is clear that the qubits can be placed in a bilayer such that in each layer one has a planar graph (with qubits on edges), while some of the checks act on qubits in both layers. Such a partitioning of the qubits can be arrived at by slicing the multi-handled surface 'through the middle'. For the Klein quartic surface we present the effect of such a slicing and the placement of qubits in Figure \ref{fig:KQbilayer}. A bilayer embedding with variable range connections could be a feasible architecture for superconducting qubits.
How to generally obtain 3D embeddings of hyperbolic surface codes which respect the symmetries of the hyperbolic tiling (such as the dodecadodecahedron) is an interesting mathematical question. A necessary requirement for this is that the group of symmetries of the lattice shares a subgroup with the orthogonal group in three dimensions. It will be interesting to study the performance of these codes in the presence of more realistic noise models as has been done for the surface code. More work remains to be done as well on constructing variations and optimizations of realizing access to storage and/or the fault-tolerant movement of qubits in storage.

\section*{Acknowledgements}
\addcontentsline{toc}{section}{Acknowledgements}
We would like to thank Kasper Duivenvoorden and Michael Kastoryano for fruitful discussions. BMT, NB and CV acknowledge support through the EU via the ERC GRANT EQEC (No. 682726), CV also acknowledges support by the Excellence Initiative of DFG and ETC is supported by the EPSRC (EP/M024261/1). AK would like to thank the IQI and RWTH Aachen University for their hospitality during his stay.

\appendix 

\section{Finding Quantum Codes Based on a $\{r,s\}$-Tiling of the Hyperbolic Plane}
\label{app:tiling}

Two-dimensional regular tilings cover surfaces by regular polygons ($r$-gons) which are all of the same type and with the same number of polygons ($s$) meeting at every vertex. Here we give a short summary of how one can construct hyperbolic code families based on such $\{r,s\}$-tiling. A more detailed description can be found in \cite{BT:hyper}. One could extend this construction to the more general class of uniform tilings (semi-regular and quasi-regular) in which different polygons are used.

Consider a $\{r,s\}$-tiling of the infinite hyperbolic plane with $1/r+1/s < 1/2$. A regular $r$-gon can be triangulated by identical right-angled triangles so that each side of a triangle represents an axis about which one can reflect the triangle to obtain another triangle. Each triangle having three sides, there are three reflections which we call $a$, $b$ and $c$.  If we have a regular tiling in which $s$ $r$-gons meet at every vertex, then let the angle between $a$ and $b$ be $\pi/2$, the angle between $b$ and $c$ be $\pi/r$ and the angle between $a$ and $c$ is $\pi/s$. The triangle is thus a hyperbolic triangle where the sum of its three angles is less than $\pi$.

Let the group of reflections generated by $a, b$ and $c$ be called $G_{r,s}$. $G_{r,s}$ is a group with a countably infinite number of elements but it is finitely generated by $a, b$ and $c$:
\begin{equation}
  G_{r,s} = \langle a,b,c \mid a^2=b^2=c^2=(ab)^2=(bc)^r=(ca)^s=e \rangle 
\label{eq:grs}
\end{equation}
where $e$ is the identity element of the group. Here $\rho=bc$, $\sigma=ca$ and $\tau=ab$ are rotations, i.e. $\rho$ acts as a clockwise $2\pi/r$ rotation around the center of a face and $\sigma$ acts as a clockwise $2\pi/s$ rotation around a vertex. The relations between $a,b,c$ in (\ref{eq:grs}) express the fact that $a,b,c$ are reflections and that $\tau,\rho$ and $\sigma$ are rotations. Starting from some triangle, any other triangle can be obtained by applying an element in $G_{r,s}$ to the initial triangle so that the group elements of $G_{r,s}$ label the triangles. From the reflections $a,b,c$ one can obtain all the symmetry transformations of the lattice, i.e. all rotations and translations. Note that our convention here is to multiply elements from the right, so $ab$ reads 'first apply $a$, then $b$'.

There is an important subgroup of $G_{r,s}$ denoted as $G_{r,s}^{+}$ which only consists of rotations and translations. This group $G_{r,s}^{+}$ is generated by $\rho$ and $\sigma$:
\begin{equation}\label{eqn:gamma_plus}
  G_{r,s}^{+} = \langle \rho, \sigma \mid \rho^{r}=\sigma^{s}=(\rho \sigma)^{2}=e \rangle .
\end{equation}
Note that the rotation $\tau \in G_{r,s}^+$.

To define a quantum code based on this tiling one has to find a torsion-free normal subgroup $H \subseteq G_{r,s}^+$ (a normal subgroup has the property that $g H= H g$ for all $g \in G_{r,s}^+$) which gives rise to the quotient group $G_H=G_{r,s}^+ / H=\{ gH: g \in G_{r,s}^+\}$. Torsion-freeness of this subgroup $H$ means that there is no element $h \in H$ such that $h^m=e$ for some finite $m$: it implies that $H$ only contains translations.  The Todd-Coxeter algorithm allows one to find all normal subgroups $H$ such that the quotient group $G_H$ has a low number of elements (called the index of $H$ in $G_{r,s}^+$) given the description of a finitely-generated group such as $G_{r,s}^+$, see \cite{CD:low_index}. The condition that the subgroup is torsion-free can then be checked by checking whether the subgroup contains elements of finite order.

Given $G_H$, one can identify the faces, vertices and edges of the resulting code lattice with the following objects. Note that $G_H$ contains the cyclic subgroups $\langle \rho\rangle$ and $\langle \sigma\rangle$.  
A face of the code lattice can then be identified with a left coset of $\langle \rho\rangle$ in $G_H$, i.e. $g \langle\rho\rangle$ with $g \in G_H$. Intuitively, it can be understood as follows. If we have a given triangle (and its partner triangle obtained by reflection in $b$) and we apply all elements in $\langle \rho\rangle$ one obtains the face in which the triangle (and its partner) is contained. Given that we identify faces $f \in F$ which are related by a translation in $H$, the distinct faces thus correspond to different cosets of $\langle \rho\rangle$ in $G_H$. 
Similarly, one can identify the vertices $v \in V$ of the code lattice as left cosets of $\langle \sigma\rangle$ in $G_H$. The edges $e \in E$ (and so the qubits) of the code lattice can be identified with the left cosets of $\langle \tau\rangle$ in $G_H$.
Note that $\langle \rho\rangle$,$\langle \sigma\rangle$ and  $\langle \tau\rangle$ are not normal in $G_H$ so that they do not form a Cayley graph.
The enumeration of these different cosets thus allows one to get a complete numerical description of the stabilizer group of the code.
Using Euler's formula for closed surfaces, i.e. $\chi=2-2g=|V|-|E|+|F|$, this allows one to determine the genus $g$ (number of handles) of the surface. The code encodes $2g$ logical qubits, two qubits per handle.

The logical operators can be constructed by defining boundary $\partial_i \colon C_i\rightarrow C_{i-1}$ and co-boundary operators $\delta_i \colon C_i\rightarrow C_{i+1}$. Here $C_i$ ($i=F,E, V$) is a $d_i$-dimensional (resp. $d_i=|F|,|E|, |F|$) $\mathbb{Z}_2$-vectorspace whose basis elements correspond to faces, edges and vertices respectively. The boundary operator $\partial_2$ maps a face (or collections of faces) onto the set of edges which form the boundary of the face (collection of faces). The co-boundary operator $\delta_0$ maps a vertex (or collection of vertices) onto a set of edges which are incident to this vertex (collection of vertices).  Similarly, one can use the boundary operator $\partial_1 \colon C_E \rightarrow C_V$ which maps edges to vertices and the co-boundary operator $\delta_1 \colon C_E \rightarrow C_F$, mapping edges to faces.

The generators of ${\rm Im}(\partial_2)$ correspond to the $Z$-checks of the code. The $\overline{Z}$ operators of the code are elements of ${\rm Ker}(\partial_1)$ (since these operators are closed loops, they have no vertex boundary) which are not contained in ${\rm Im}(\partial_2)$.  Similarly, the generators of ${\rm Im}(\delta_0)$ correspond to the $X$-checks of the code. The $\overline{X}$ operators are elements of ${\rm Ker}( \delta_1)$ which are not contained in ${\rm Im}(\delta_0)$. Using linear algebra one can thus obtain a basis of $\overline{Z}$ and $\overline{X}$ operators, i.e. a set of mutually commuting pairs $\overline{X}_i, \overline{Z}_i, i=1, \ldots, k$. In \ref{sec:dist_comput} we discuss how one can efficiently determine the distance of the code and compute the number of logical operators of this minimal weight.

\section{Efficient Computation of Distance for Any CSS Surface Code}
\label{sec:dist_comput}

We describe an algorithm formulated by Bravyi \cite{bravyi_algo} which allows one to efficiently compute the distance of any CSS surface code encoding $k$ logical qubits into $n$ qubits. It also allows one to count the number of $\overline{Z}$ (or $\overline{X}$) operators of this minimum weight which we need in Section \ref{sub:approximation} in the approximate formula for the logical error probability.

Since the code is a CSS code, the minimum weight logical operator is an operator which is either only $X$-like or $Z$-like. We focus on calculating $d(\overline{Z})$, i.e. the minimum weight of $\overline{Z}$, but the procedure for $d(\overline{X})$ is identical.

Let $G=(V,E)$ be the graph associated with the code.  Take a $\overline{X}$ operator, say $\overline{X}_1$ and let $E(\overline{X}_1)\subseteq E$ be its qubit support. Take two copies of the graph, $G$ and $G'$. Using these copies, we define a new graph $\tilde{G}=(\tilde{V},\tilde{E})$ with $\tilde{V}=V \cup V'$ and the following edge set $\tilde{E}$.  Omit in $\tilde{E}$ each edge $e=(u,v) \in E(\overline{X}_1)$ and $e'=(u',v') \in E(\overline{X}_1)$. For these omitted edges we instead include two new {\em cross-over} edges $(u,v') \in \tilde{E}$ and $(u',v) \in \tilde{E}$.  For all other edges $e \in E-E(\overline{X}_1), e'\in E-E(\overline{X}_1)$, include $e$ and $e'$  in $\tilde{E}$.

Consider the shortest graph distance $d(v, v')$ in $\tilde{G}$. Since $v\in G$ and $v' \in G'$, any path ${\cal P}$ from $v$ to $v'$ has to cross over an odd number of times from $G$ to $G'$ or vice versa. The path ${\cal P}$ can thus be mapped to a loop in the graph $G$ which has an odd overlap with the support of $\overline{X}_1$:  we start at vertex $v$ and we replace each cross-over edge $(u',v)$ or $(u,v')$ that we encounter on the path ${\cal P}$ by the orginal edge $(u,v)$. We obtain a path ${\cal P}$ that will stay in the graph $G$ and which comes back to $v$ itself. Since the number of cross-over edges used is odd, this closed $\overline{Z}$ loop in $G$ will anti-commute with $\overline{X}_1$. 

Thus in order to determine the minimum weight of a $\overline{Z}$ operator which anti-commutes with $\overline{X}_1$, we iterate over the points $v$ such that $(u,v) \in E(\overline{X}_1)$, and for each choice of $v$ one determines the shortest graph distance $d(v,v')$. The shortest graph distance is calculated using Dijkstra's algorithm which is efficient in the number of vertices of the graph. One then takes the minimum over all these graph distances. Of course the found $\overline{Z}$ may also anti-commute with other $\overline{X}_i$.

In order to determine $d(\overline{Z})$ we iterate over the elements of the logical operator basis $\overline{X}_1, \ldots, \overline{X_k}$.
Assuming a list of $\overline{X}_1,\ldots, \overline{X}_k$, the procedure is $O(k n^2 \log n)$ where $n=|E|$.\\

To determine the number of $\overline{Z}$ operators of minimum weight $d(\overline{Z})$ (called $N^Z_d$), one has to be slightly more careful to avoid double counting. Given a fixed $\overline{X}_i$ and a fixed vertex $v$, one can run Dijkstra's algorithm to determine all paths ${\cal P}$ between vertices $v$ and $v'$ of given minimal distance. One then just collects these paths into a list and keeps adding to the list, avoiding doublecounts, by iterating over $v$ and then over all $\overline{X}_i$. The results on $N^Z_d$ and $N^X_d$ are shown in Table \ref{tab:num_short_logicals}.

We have seen in Section \ref{sec:HSC} that the hyperbolic code with 60 qubits can be identified with a self-intersecting polytope called the dodecadodecahedron. Looking at this Figure and its symmetries one can understand the finding that there are 30 lowest-weight $\overline{Z}$ operators and 90 lowest-weight $\overline{X}$ operators. In Figure \ref{fig:dodecadodecahedron} (b) we see that some weight-6 $\overline{X}$ are in one-to-one correspondence to the set of vertices (a vertex is highlighted in black). Hence there are at least 30 $\overline{X}$ of length 6.
In Figure \ref{fig:dodecadodecahedron} (c) we see another type of weight-6 $\overline{X}$ which is incident to two edges of two yellow faces.
Rotating one of those faces gives a new logical operator.
Since there are two yellow faces per logical operator we will over count by a factor of 2.
Hence the number of these type of $\overline{X}$  is $|F|\times 5/2 = 60$, thus 90 all together.
Similarly, from Figure \ref{fig:dodecadodecahedron} (d) we see that there must be at least 30 $\overline{Z}$ of weight 4, which is the number computed using the algorithm on this code.

\begin{figure}
    \begin{center}
    \includegraphics[width=.45\textwidth]{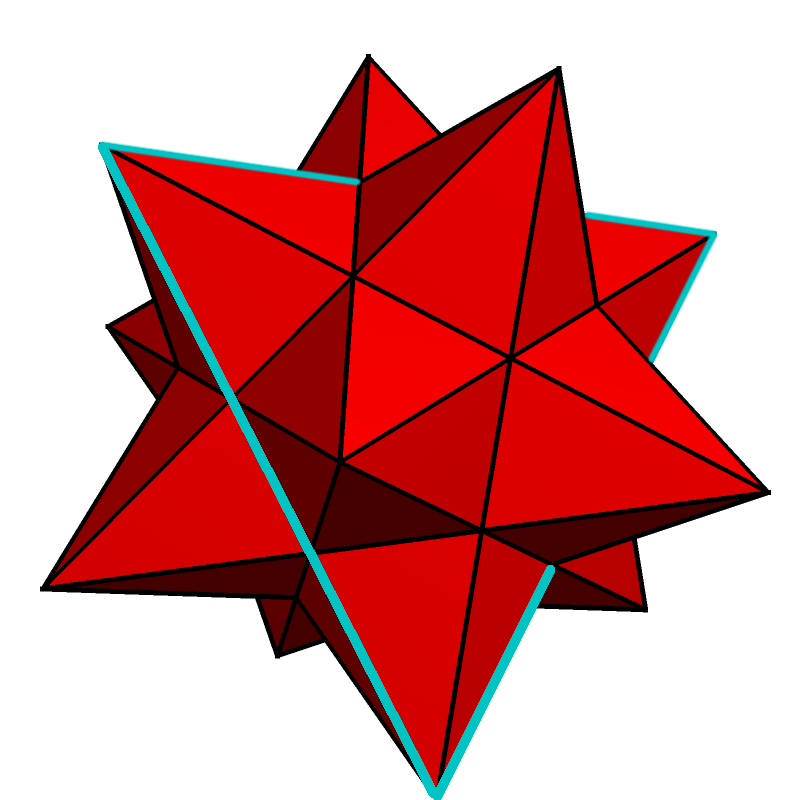}
    \end{center}
    \caption{The small stellated dodecahedron with 12 vertices, 30 edges and 12 faces. Highlighted in blue is a $\overline{Z}$ of weight 3. Every vertex has 5 different such loops running through it. Since we are overcounting by a factor of three there are $5\times 12 / 3 = 20$ such loops. This is exactly the result of the algorithm.}
    \label{fig:small_stellated_dodecahedron}
\end{figure}

Another such example is a $\{5,5\}$-tiling of a surface with the same genus 4, instead of a $\{5,4\}$-tiling.
It corresponds to a $[[30,8,3]]$ quantum code with 12 $X$- and 12 $Z$-checks, both of weight 5.
It can be found in Table III in \cite{BT:hyper}.
Similarly as in the previous case we can turn the pentagons into self-intersecting pentagrams and arrange the vertices on the surface of a sphere to obtain the {\em small stellated dodecahedron} (see Figure \ref{fig:small_stellated_dodecahedron}).

\begin{table}[htb]
	\begin{center}
        \begin{tabular}{|c|c|c|c|}
            \hline $n_h$ & $l$ & $N_d^{Z}$ & $N_d^{X}$ \\
            \hline 60 & 1 & 30 & 90 \\
            \hline 60 & 2 & 30 & 60 \\
            \hline 60 & 3 & 30 & 60 \\
            \hline 60 & 4 & 30 & 60 \\
            \hline 60 & 5 & 30 & 60 \\
            \hline 60 & 10 & 30 & 60 \\
            \hline 160 & 1 & 320 & 500 \\
            \hline 160 & 2 & 2880 & 6560 \\
            \hline 160 & 3 & 32000 & 93760 \\
            \hline 360 & 1 & 5670 & 90 \\
            \hline 1800 & 1 & 31320 & 180 \\
            \hline
		\end{tabular}
	\end{center}
    \caption{The number of minimum-weight $\overline{Z}$ and $\overline{X}$ operators for hyperbolic and semi-hyperbolic surface codes based on the $\{4,5\}$-tiling. Note that for the semi-hyperbolic lattices with $n_h=60$ the number of minimum-weight logical operators, curiously, does not increase. This means that $\overline{Z}$ never enters and leaves any vertex by two adjacent edges in the hyperbolic code $[[60,8,4]]$. If the string were to enter and leave a vertex by adjacent edges, thus touching a face in at least two edges, then upon subdivision of the face by $l$, there will be ${2l \choose l}$ ways of rerouting the string while keeping its length the same. This fact can be verified by considering the logical $Z$ operators in Figure \ref{fig:dodecadodecahedron}(c). The blow-up in the number of logical operators that we see for $n=160$ might relate to such rerouting of strings, but we have not investigated this in detail.}
    \label{tab:num_short_logicals}
\end{table}

\section*{References}
\addcontentsline{toc}{section}{References}
\bibliographystyle{hunsrt}
\bibliography{hyperbolic}

\end{document}